\documentclass[sigconf]{acmart}%\documentclass[sigconf]{acmart}
\AtBeginDocument{%
  }

\copyrightyear{2026}
\acmYear{2026}
\setcopyright{cc}
\setcctype{by}
\acmConference[NordiCHI '26]{Proceedings of the 14th Nordic Conference on Human-Computer Interaction}{October 03--07, 2026}{Vaasa, Finland}
\acmBooktitle{Proceedings of the 14th Nordic Conference on Human-Computer Interaction (NordiCHI '26), October 03--07, 2026, Vaasa, Finland}
\acmDOI{10.1145/3829807.3829882}
\acmISBN{979-8-4007-2373-5/2026/10}

\usepackage{caption}
\usepackage{subcaption}
\usepackage{tabularx} % should be fine?
\usepackage{multirow}
\usepackage{colortbl}
\usepackage{url} %strange: the URLs still do not work...

\begin{document}

\title{Exploratory Unstructured Data Analysis: A Formative Study and Implications for Human-AI Collaboration}

\author{Johannes Eschner}
\orcid{0009-0001-6784-8503}
\affiliation{
  \institution{TU Wien}
  \city{Vienna}
  \country{Austria}
}
\email{jeschner@cg.tuwien.ac.at}

\author{Dominik Eitler}
\orcid{0009-0006-6417-938X}
\affiliation{
  \institution{TU Wien}
  \city{Vienna}
  \country{Austria}
}
\email{dominik.eitler@tuwien.ac.at}

\author{Max Irendorfer}
\orcid{0009-0001-0147-6954}
\affiliation{
  \institution{TU Wien}
  \city{Vienna}
  \country{Austria}
}
\email{max.irendorfer@tuwien.ac.at}

\author{Patrick Kramml}
\orcid{0009-0007-2984-9461}
\affiliation{
  \institution{University of Applied Sciences St. Pölten (USTP)}
  \city{St. Pölten}
  \country{Austria}
}
\email{patrick.kramml@ustp.at}

\author{Matthias Zeppelzauer}
\orcid{0000-0003-0413-4746}
\affiliation{
  \institution{University of Applied Sciences St. Pölten (USTP)}
  \city{St. Pölten}
  \country{Austria}
}
\email{matthias.zeppelzauer@ustp.at}

\author{Manuela Waldner}
\orcid{0000-0003-1387-5132}
\affiliation{
  \institution{TU Wien}
  \city{Vienna}
  \country{Austria}
}
\email{waldner@cg.tuwien.ac.at}

\renewcommand{\shortauthors}{Eschner et al.}

\begin{abstract}    
    We propose a conceptual framework for exploratory data analysis of (large) unstructured data (EluDA), combining classical elements (querying, visualization) with active knowledge construction in the ``search for structure''. In a formative study, users conceptualized a structure for an image dataset during exploration. We found that users conceptualize by building faceted classifications bottom-up and rarely create meaningful spatial categorization during this process. We also evaluated CLIP for zero-shot assignment and semantic categorization, finding that it remains unreliable for assigning user-defined concepts to images but does support semantic grouping. Based on these findings, we identify and discuss four key opportunities for human-AI collaboration in EluDA: intelligent sampling and visualization to maximize data visibility; incremental and few-shot learning to minimize effort for reliable assignment; automatic category, concept, and facet suggestions to reduce effort during the search for structure; and the necessity for effective trust calibration methods.
\end{abstract}

\begin{CCSXML}
<ccs2012>
   <concept>
       <concept_id>10003120.10003121.10011748</concept_id>
       <concept_desc>Human-centered computing~Empirical studies in HCI</concept_desc>
       <concept_significance>500</concept_significance>
       </concept>
   <concept>
       <concept_id>10003120.10003145.10011770</concept_id>
       <concept_desc>Human-centered computing~Visualization design and evaluation methods</concept_desc>
       <concept_significance>300</concept_significance>
       </concept>
   <concept>
       <concept_id>10010147.10010178.10010187</concept_id>
       <concept_desc>Computing methodologies~Knowledge representation and reasoning</concept_desc>
       <concept_significance>100</concept_significance>
       </concept>
   <concept>
       <concept_id>10003120.10003145.10003147.10010365</concept_id>
       <concept_desc>Human-centered computing~Visual analytics</concept_desc>
       <concept_significance>300</concept_significance>
       </concept>
 </ccs2012>
\end{CCSXML}

\ccsdesc[500]{Human-centered computing~Empirical studies in HCI}
\ccsdesc[300]{Human-centered computing~Visualization design and evaluation methods}
\ccsdesc[300]{Human-centered computing~Visual analytics}
\ccsdesc[100]{Computing methodologies~Knowledge representation and reasoning}

\keywords{Exploratory data analysis, visual analytics, mixed-methods evaluation}
\begin{teaserfigure}
\centering
\includegraphics[width=\textwidth]{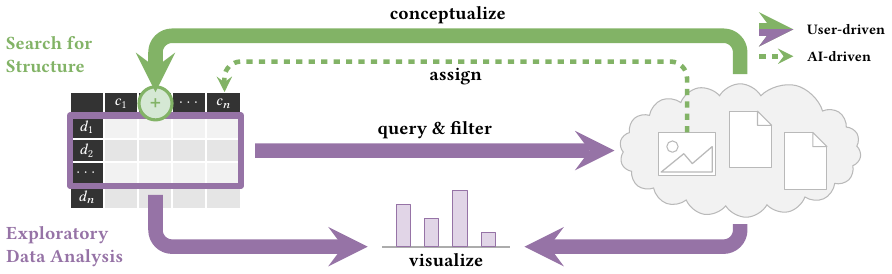}
\caption{We frame exploratory analysis of (large) unstructured data (EluDA) as an interplay between classic exploratory data analysis (EDA), which requires well-structured data, and an incremental data-, user- and AI-driven search for structure. Purple elements signify EDA processes, while green represents the ``search for structure'' in unstructured data analysis. Two arrow styles represent two levels of agency: regular arrows indicate user-driven, AI-supported tasks, while the dashed arrow in the assign step represents the proposed AI-driven data assignment.}
\Description{Diagram showing the interaction between exploratory analysis of structured data (left) and the search for structure in unstructured data (right). On the left is a small table representing structured data, with dark header cells and one highlighted block. On the right is a cloud containing unstructured documents (an image icon and two file icons). Between them are thick directional arrows and labels. A thick green arrow labeled ``conceptualize'' runs across the top between the cloud side and the table side; a plus on the table side indicates the addition of a new column. The middle path is a thick purple arrow labeled ``query \& filter'' running from the table toward the cloud. A dashed green path labeled ``assign'' runs from a document in the cloud back toward the table. The bottom path labeled ``visualize'' consists of two thick purple arrows pointing inward toward a small bar chart at the bottom center. Section labels appear on the left: ``Search for Structure'' (green, upper left) and ``Exploratory Data Analysis'' (purple, lower left). A legend in the top-right explains arrow types: a solid arrow means user-driven, and a dashed arrow means AI-driven.}
\label{fig:teaser}
\end{teaserfigure}


\maketitle

\section{Introduction}
 Exploratory data analysis (EDA)~\cite{tukey_exploratory_1977,hartwig_exploratory_1979} is a process for analyzing datasets aimed at uncovering their main characteristics, particularly in situations where both the relevant patterns and the target of the analysis are initially unknown. Usually, EDA is based on summary statistics and relies heavily on visual displays of data, such as histograms and scatterplots, as well as interaction techniques like querying and filtering. Traditional EDA methods are thereby designed for structured data, i.e., data organized in a well-defined tabular form with semantically meaningful attributes, in relational databases, or in semi-structured representations such as XML or JSON~\cite{young_visual_2006, theus_interactive_2008}. For unstructured data sources such as images, videos, or document collections, traditional analysis methods do not apply~\cite{blumberg_problem_2003}. For example, in wildlife observation camera traps continuously collect millions of images or videos, whose semantic attributes, such as species and observed behavior, must first be identified before downstream analysis can be performed, making annotation a major bottleneck~\cite{gabeff_wildclip_2024}. Therefore, a first step prior to analysis often involves deriving the unstructured data into a new, structured form, for example, a table~\cite{wongsuphasawat_goals_2019}. This process is not well supported by current tools, which follow an ``attribute-centric approach''~\cite{stolte_polaris_2002,tableau}, it is not well understood, and is assumed to be especially challenging when dealing with very large unstructured data. Given that unstructured data is estimated to represent up to 80\% of all data~\cite{li_tetrahedral_2010}, developing strategies for the so-called ``search for structure'' is therefore of high interest~\cite{buja_statistical_2009, stolte_polaris_2002}. In the context of \textit{data lakes} (i.e., repositories storing large amounts of raw, heterogeneous data~\cite{hai_data_2023}), dataset organization and structuring for subsequent exploration have been identified as an essential part of the analysis~\cite{nargesian_organizing_2020, hai_data_2023}.

To this end, we propose a conceptual framework of exploratory large unstructured data analysis (EluDA, see Figure~\ref{fig:teaser}), which combines classical elements of EDA (querying, filtering, and visualization) with active knowledge construction in the ``search for structure''. In EluDA, we distinguish two fundamental operations: (1) \emph{conceptualization}, which refers to the incremental definition of a structured description of the data based on data observations and user expectations, and (2) \emph{assignment}, which is the association of the data items to the conceptualized structure.

Assigning data items to a structure can be considered a classification problem, which has been extensively studied by many different domains~\cite{fisher_use_1936, krizhevsky_imagenet_2012, he_deep_2016, kowsari_text_2019}. In the context of interactive data analysis, approaches that actively integrate the user into the classification process are active learning~\cite{jia_towards_2022} and visual interactive labeling~\cite{bernard_vial_2018}. More recently, zero-shot classification with pre-trained foundation models, such as vision-language models (VLMs)~\cite{radford_learning_2021}, has shown promising performance for fast, approximate classification~\cite{cooper_rethinking_2025} -- also in the context of interactive data analysis~\cite{eschner_interactive_2025}. Today, VLMs have the potential to efficiently and accurately match visual content to natural language descriptions, thereby serving as a powerful foundation for analyzing large, unstructured data. 
However, classification problems typically work top-down, i.e., they assume that a (often flat) class structure already exists a priori. Conceptualization in the context of interactive data exploration -- i.e., how humans gradually discover and express concepts that describe the observed data in natural language or other forms -- has received much less attention so far. Predominantly, unstructured data analysis builds upon existing knowledge bases~\cite{wagner_kavagait_2019,li_incorporation_2022,wang_kmtlabeler_2024}. 
In practice, an appropriate knowledge base may not be available, may not be directly applicable to the present data and task, or may simply not align with the user's prior knowledge, interests, and expectations~\cite{nonaka_knowledge-creating_1998}. 

The aim of our work is therefore to deepen the understanding of how users externalize and organize their \emph{knowledge} during their ``search for structure'' in unstructured data during exploratory analysis. This understanding is essential for designing suitable data exploration methods and for assessing whether and how state-of-the-art multimodal models can eventually help us realize the EluDA loop depicted in Figure~\ref{fig:teaser}. 
The core of our work is a formative mixed-methods study with two low-fidelity conceptualization interfaces that support a variety of knowledge organization approaches. In the study, 20 participants explored a dataset of 100 images and were asked to identify and express concepts that describe the images' content. We tracked both \emph{implicit conceptualization} through users' spatial organization of images on a large multi-touch table and \emph{explicit conceptualization} via a concept map-like interface. 
Two independent coders curated user conceptualizations, and 146 crowd workers performed the assignment step indicated in Figure~\ref{fig:teaser} to simulate a fully functional (and potentially AI-assisted) EluDA cycle. We contribute three core pieces of foundational knowledge informing the design of future EluDA systems:  
\begin{enumerate}
    \item By characterizing the observed knowledge organization processes and schematic structure of users' conceptualizations, we present design recommendations for conceptualization interfaces in the EluDA process. 
    \item By using multimodal foundation models for assignment and categorization and comparing their output to human-curated ground truth assignments and user-generated categorizations, we discuss the capabilities and limitations of state-of-the-art models to support the ``search for structure''. 
    \item Based on our conceptual EluDA framework and the findings in our study, we discuss human-AI collaboration in general and multimodal models like VLMs~\cite{radford_learning_2021} in particular as key enabling technologies for EluDA systems that learn with the users and support their analysis. 
\end{enumerate}

\noindent
The curated study data is available on OSF: \url{https://osf.io/xhdmv}.

\section{Background and Related Work}

As illustrated in Figure~\ref{fig:teaser}, the gradually conceptualized structure serves as the backbone for the EluDA process. This structure can be seen as a form of \emph{explicit knowledge} ($K^\epsilon$) provided by the user. Explicit knowledge is something that is known (about the data) and can be written down in textual form~\cite{hogan_knowledge_2022}. In contrast, \emph{tacit knowledge} ($K^\tau$) resides in the user's mind~\cite{nonaka_dynamic_1994}. \emph{Knowledge externalization} is the process of transforming intangible tacit into explicit knowledge. The process of conceptualization in Figure~\ref{fig:teaser}, hence, is a form of knowledge externalization, in which users formalize, in words or other forms, what they already know about the data and potentially what they still expect. We therefore first discuss the basic principles of knowledge organization, then review knowledge externalization strategies, and finally present examples of how explicit knowledge can be exploited to facilitate (exploratory) data analysis. 

\subsection{Knowledge Organization}
\label{sec:knowledge-organization}

A natural way to make sense of the world is to \emph{classify} what we see, know, and experience~\cite{broughton_essential_2015}. 
Explicit knowledge can be organized in different structures -- from tree-like taxonomies~\cite{broughton_essential_2015}, over (hierarchical) lists and tables~\cite{dawson_need_2006} to knowledge graphs~\cite{krotzsch_editorial_2016}. These structures vary in their rigidity and expressive power: taxonomies enforce hierarchy and completeness, while graphs enable inference and reasoning and allow for more flexibility~\cite{chen_review_2020}. Well-known examples of knowledge graphs include WordNet~\cite{fellbaum_wordnet_2010} and Wikidata~\cite{vrandecic_wikidata_2014}. 
Traditional classification approaches work \emph{top-down}, i.e., individual observations are assigned to a pre-defined explicit knowledge structure. Faceted classification~\cite{jacob_classification_2004, mills_faceted_2004}, on the other hand, works \emph{bottom-up} based on concepts that have been observed from the data~\cite{prieto-diaz_faceted_2003}, which are then organized into classes (facets). The data can then be uniquely described by a compound of concepts from different facets. 
Faceted classification has been commonly used in classic library systems~\cite{hjorland_facet_2013, ranganathan_colon_1933}. 
Similar to faceted classification, tagging is a bottom-up classification process in which users apply short, natural-language descriptions (``tags'') from an open vocabulary to data items. In contrast to faceted classification, it is less formal, as it usually does not require users to organize their tags into semantic categories. Tagging has been -- and often still is -- a promising method for efficient retrieval and sharing of web resources~\cite{gupta_survey_2010} and has been frequently conducted by millions of users producing so-called ``folksonomies''~\cite{peters_folksonomies_2009}, which formed a rich repository for knowledge organization research. A common observation, for instance, is that a small number of tags are used very frequently, while most are used infrequently~\cite{halpin_complex_2007}.  

In contrast to classification, \emph{categorization} describes the process of dividing the world into groups of entities whose members are considered equivalent~\cite{rosch_basic_1976,jacob_classification_2004}. Cognitive psychology has intensively studied how humans form categories. Rosch~\cite{rosch_basic_1976}, for instance, argues that a category system has to provide maximum information with the least cognitive effort. She further argues that this can be achieved by mapping categories to known attribute structures.
In a series of experiments, she found that humans form categories at a \emph{basic} level. Basic categories occupy a middle ground between the abstract and the concrete, allowing for a representative mental image of all category members. 

While classification, tagging, and categorization have been intensively studied in cognitive psychology and information retrieval, knowledge organization during EDA has received much less attention so far. It is unclear whether analysts aim to first construct an explicit knowledge structure to later classify the observed data top-down, whether they rather observe the data first and then derive a set of tags or faceted concepts therefrom in a bottom-up manner, or whether they prefer to categorize the data into equivalent groups without a natural-language description thereof, while potentially deriving a taxonomy or set of tags from these categories. These are fundamentally different strategies, which therefore require distinct knowledge externalization interfaces. It is also an open question how to ultimately integrate AI assistance for more efficient data analysis while leaving the user in control.

\subsection{Knowledge Externalization}
\label{sec:knowledge-externalization}

The process of externalizing knowledge has long been established in social sciences, serving communication, training, and documentation~\cite{klein_critical_1989, gavrilova_knowledge_2012}. In the context of Visual Analytics (VA), the interactive externalization of tacit knowledge~\cite{nonaka_knowledge-creating_1998} has been proposed as a core component of the analysis process~\cite{wang_defining_2009}. Within the framework of Knowledge-Assisted Visual Analytics~\cite{federico_role_2017}, there are two paths for transferring tacit knowledge into machine-readable explicit knowledge: 1) explicit representation in a \emph{direct externalization interface} or 2) implicit inference from users' interactive data exploration through \emph{interaction mining}.

Explicit externalizations support classification-like knowledge organization. They can be used in a top-down manner, for instance, by showing an existing knowledge graph or ontology~\cite{lohfink_knowledge_2022}, or can be constructed interactively by the user in a bottom-up fashion while exploring the data. Bottom-up construction of explicit knowledge externalizations can take multiple forms -- from informal textual notes to strict hierarchical representations, such as a mind map, to graph-based representations, such as a concept map~\cite{novak_theory_2006}. Originally from the educational domain, concept maps describe a given topic through concepts (nouns) and their relations (verbs). In a concept map, more general concepts are usually arranged at the top, and more specific concepts at the bottom. Concept maps are generally considered intuitive in their construction~\cite{watson_assessing_2016, starr_concept_2013}. In the context of EDA, bottom-up graph-based knowledge representations have been used, for instance, for knowledge transfer between analysts~\cite{zhao_supporting_2018, mahyar_closer_2010}, as well as for personal structured note-taking~\cite{waldner_linking_2021}. Limitations are that they can become very large~\cite{zhao_supporting_2018, mahyar_supporting_2014}, are considered effortful to create and maintain~\cite{zhao_supporting_2018}, and often result in a rather casual structure, which may be difficult to parse~\cite{waldner_linking_2021}.

Implicit knowledge externalization interfaces, such as the spatial organization of data items on a large display, support categorization approaches. Spatialization simplifies perception, as larger groups with shared features are easier to keep track of than individual items~\cite{kirsh_intelligent_1995}. People create spatial organization of documents on a table using piles~\cite{malone_how_1983, mander_pile_1992} and prefer spatial organization over strict filing. Spatialization also applies to large vertical displays, where users create spatial document clusters~\cite{andrews_space_2010}, often reflecting topical similarities~\cite{endert_semantic_2012}. However, spatial clustering of data items is highly subjective~\cite{cetin_visual_2018} and is limited by constraints on available physical space and the lack of explicit textual labels. Also, when given the option to explicitly externalize observations in a graph-based structure, users mostly refrained from spatially organizing data items~\cite{waldner_linking_2021}. 

To get a more holistic understanding of how analysts organize their knowledge when exploring unstructured data, we therefore provided users with low-fidelity interfaces that support explicit externalization for classification-like knowledge organization and implicit externalization for data categorization by spatial organization. 

\subsection{Knowledge Exploitation}
\label{sec:knowledge-exploitation}

Already decades ago, \emph{knowledge-based systems}~\cite{buchanan_dendral_1981} exploited explicit knowledge to derive insights via automated reasoning from data. Nowadays, large knowledge bases, such as curated knowledge graphs, are foundational to AI systems, including recommender systems, LLMs, and question-answering systems~\cite{peng_knowledge_2023}. In visual analytics, explicit knowledge is exploited to guide visualization, interaction, and collaboration, for instance, through an explicit ``knowledge store''~\cite{wagner_kavagait_2019}, or an ``acting ontology''~\cite{lohfink_knowledge_2022} for classification. Others use user-defined classes to steer embeddings~\cite{li_incorporation_2022} in a bottom-up manner, support active learning~\cite{wang_kmtlabeler_2024}, or guide exploration through user feedback~\cite {ge_exnav_2020}. 

EDA can also be supported by categorization-like knowledge organization. \emph{Semantic interaction}~\cite{endert_semantic_2012}, for example, learns semantic similarities between documents and the importance of terms from user interactions, such as categorization through spatial organization of documents or text highlighting. Similarly, Guo et al.~\cite{guo_expert---loop_2016} update a topic model based on users' image re-arrangements to gradually improve semantic image groupings. 

Recent advances in natural language interaction with AI systems have opened new opportunities for human-AI collaboration in data analysis~\cite{setlur_supporting_2025} -- particularly in combination with knowledge organization approaches that rely on natural language descriptions, such as classification or tagging. %In our work, we investigate how users intuitively organize and externalize their knowledge when exploring unstructured data and discuss how this knowledge can be exploited for human-AI collaboration to support exploratory analysis in the future. 
Although our study does not evaluate such an AI system in its entirety, understanding how users organize and externalize their knowledge during data exploration is a prerequisite for designing effective human-AI collaboration.

\section{Formative Study}

The aim of our formative, exploratory study was to build an initial understanding of how users gradually conceptualize a structure describing an unstructured, previously unknown dataset. More specifically, we aimed to answer the following research questions with respect to the EluDA framework (Figure~\ref{fig:teaser}) with our study: 

\begin{description}
    \item[\textbf{RQ1:}] Which knowledge organization strategies do users apply for conceptualization during data exploration?  
    \item[\textbf{RQ2:}] How reliably can state-of-the-art multimodal models support the ``search for structure''?
    \item[\textbf{RQ3:}] How useful are users' conceptualizations for EDA?
\end{description}

In our study, the unstructured dataset consisted of 100 images with a generally understandable overall theme -- generated images of persons working in different occupations. Participants were instructed to explore the variability in the visual outputs of the generative image model. Specifically, users were asked to report any concepts that were similar or dissimilar between the generated images depicting people working in different occupations (beyond the occupations themselves). The study was conducted in a lab using two low-fidelity conceptualization interfaces, designed to support a wide range of knowledge organization approaches (Section~\ref{sec:interfaces}) to investigate \textbf{RQ1}. Furthermore, two independent coders extracted a set of concepts expressed by users, and we used crowdsourcing to create a ground-truth assignment linking data items to users' conceptualizations. We then evaluated how well a state-of-the-art VLM can predict image-to-concept assignments and categories (Section~\ref{sec:assignment}), as formulated in \textbf{RQ2}. We also theoretically assessed whether users' conceptualized structures are sufficient to support basic EDA operations, such as querying and filtering (\textbf{RQ3}, Section~\ref{sec:EDA}). Based on our findings, we derive a set of guidelines for conceptualization interfaces for EluDA and discuss opportunities for machine-assisted ``search for structure'' and exploratory analysis in the EluDA process. 

\subsection{Conceptualization Interfaces (RQ1)}
\label{sec:interfaces}

\begin{figure}
  \centering
  \includegraphics[width=\columnwidth]{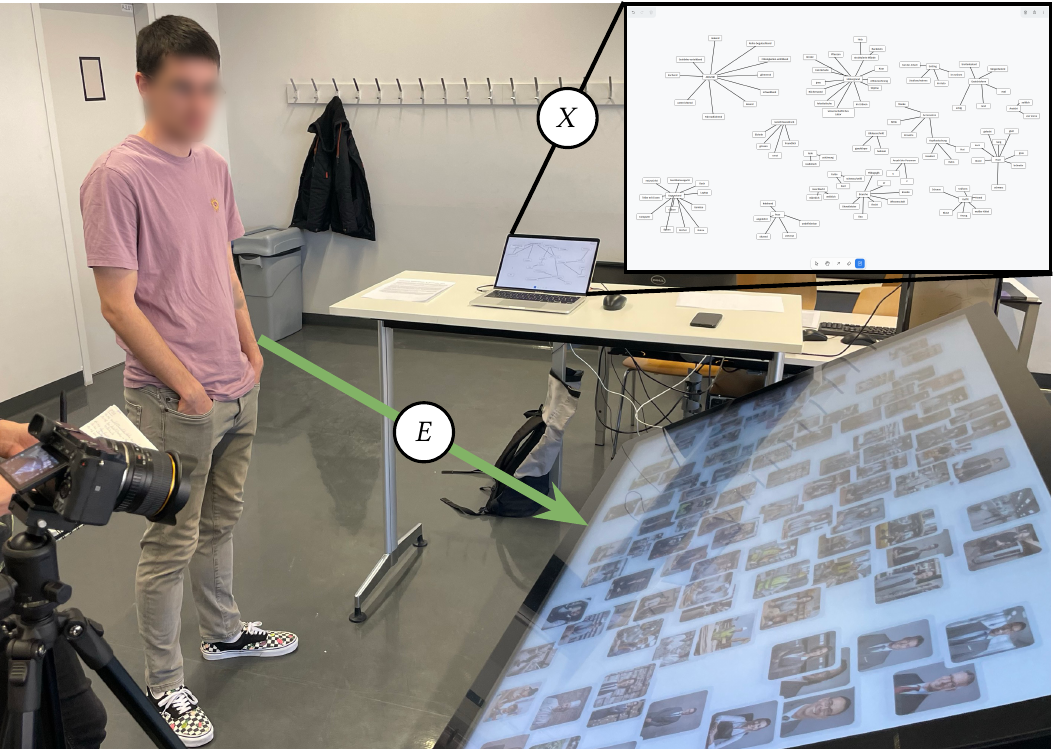}
  \caption{Study setup: The exploration interface ($E$) is provided by the large multi-touch table, and the explicit conceptualization interface ($X$) is provided on the laptop (magnified in the top right).}
  \label{fig:study_setup}
  \Description{Photo of the study setup showing a participant looking at the multi-touch table, which displays the implicit conceptualization interface. An arrow, labeled E, points from the participant's hand to the touch table. In the background, a laptop with the explicit conceptualization interface is situated on a table. In the top-right corner of the image, an enlarged view of the explicit interface on the laptop is inserted. This interface is labeled X.}
\end{figure}

Following the conceptual KAVA framework~\cite{federico_role_2017} (see Section~\ref{sec:knowledge-externalization}), we provided two low-fidelity interfaces for conceptualization: one for implicit conceptualization through spatial organization of the images while exploring them, and one for explicit conceptualization through natural language descriptions of observed concepts and -- optionally -- relations between these concepts. Figure~\ref{fig:study_setup} shows the study setup with the two independent conceptualization interfaces. We tracked all user interactions in both interfaces in 10-second intervals by persisting the interface state at every timestep. 

\paragraph{Implicit Conceptualization Interface}
\label{sec:interface_E}

We presented the images on a multi-touch table, providing space for spatial organization as a form of implicit categorization. The assumption was that the proximity of images indicates semantic similarity, relations, or hierarchy, and that it is possible to derive categories of similar images. 
The interface (see Figure~\ref{fig:study_setup}~$E$) allowed users to freely move and rotate images and restack them. Thus, it resembled the interaction with physical items on a table. Using this virtual representation allowed for tracking whether a user had uncovered an image, its location, and whether it was occluded by another image.

\paragraph{Explicit Conceptualization Interface}
\label{sec:interface_X}

The explicit conceptualization interface allows users to express concepts describing the data through natural language and relations. Concepts are represented by text boxes arbitrarily positioned and optionally connected by (labeled) links representing hierarchical organization or other relations between concepts. With this interface, it is possible to create graphs such as concept maps, mind maps, tabular arrangements of concepts, or implicitly categorized concepts based on proximity. In the following, we refer to the structure as ``concept maps'', acknowledging that our interface intentionally did not enforce rules for concept mapping. In the study setup (see Figure~\ref{fig:study_setup}~$X$), the explicit conceptualization interface was provided on a device separate from the multi-touch display.

\subsection{Machine-Assisted ``Search for Structure'' (RQ2)}
\label{sec:assignment}

We simulated two scenarios in which multimodal models support the user's ``search for structure'' during EluDA: In the first scenario, the multimodal model performs fully autonomous assignment of user-expressed concepts to images. In the second scenario, we use the model's embedding space to autonomously find and show categories of semantically similar images to the user. All analyses for RQ2 were conducted a posteriori, i.e., after data collection from the users. 

\paragraph{Scenario 1: Zero-Shot Assignment}
In an ideal implementation of the EluDA framework, assignment of data items to concepts occurs on-the-fly during conceptualization, with minimal or no user intervention, and without introducing latency in the user's exploration process. 
Multimodal models embed data from several related modalities in a common embedding space. This enables us to efficiently compute the similarity between a textual concept label externalized in the concept map interface with a data item of a different data modality -- in our case, images. Given the $j$'th data item $d_j$ and the $k$'th concept externalized by the user $c_k$, we first need to find their embedding vectors $e(d_j)$ and $e(c_k)$ in the joint space. For our study, we used the pre-trained image and text encoders of the widely used CLIP~\cite{radford_learning_2021} (openai/clip-vit-base-patch32) model to embed all images and concept labels in the users' concept maps into a joint 512-dimensional embedding space. The similarity between $d_j$ and $c_k$ can be calculated by the normalized cosine similarity between their embedding vectors: 
\begin{equation}
    sim(d_j,c_k) = \frac{e({d_j}) \cdot e({c_k}) + 1}{2}, 
    \label{eq:clip-sim}
\end{equation}
resulting in a similarity score between $0$ and $1$ per pair.
Using optimized CLIP models, computing the embedding vector and the similarity can be done in a few milliseconds, even on resource-constrained devices~\cite{vasu_mobileclip_2024, zhong_clip4retrofit_2025, yang_mobileviclip_2025}. While this is definitely fast enough for an interactive, real-time system, the question remains whether CLIP scores are sufficient to match user-defined concepts to the images (i.e., whether CLIP reliably assigns concepts to images).

\paragraph{Scenario 2: Automatic Semantic Categorization}
A popular method for visualizing unstructured data is to compute a low-dimensional (typically 2D) projection of the high-dimensional embedding vectors representing the data items using dimensionality reduction. We use UMAP~\cite{mcinnes_umap_2018} to compute such a two-dimensional projection from the CLIP image embeddings $e(d_j)$ as it represents the global structure well while also preserving local neighborhoods~\cite{espadoto_toward_2021,becht_dimensionality_2019}. If automatic categorization, using CLIP and UMAP, is able to reveal concepts relevant to users, it would be a good initial overview. Such spatialized groups are easier to keep track of than individual items~\cite{kirsh_intelligent_1995}.

\subsection{Exploratory Data Analysis (RQ3)}
\label{sec:EDA}

EDA processes such as filtering and querying only work reliably if the data are sufficiently described by the conceptualization and if all data items have been correctly assigned to the concepts. Since our study was formative and did not include essential parts of the EluDA framework, such as the assignment, we did not provide any EDA features to the user. However, we used the ground-truth assignment to analyze a posteriori whether users' conceptualizations are detailed enough to describe every image with at least one concept, and how many images can be uniquely described by multiple concepts. This enables us to estimate the coverage and exhaustiveness of the users' externalizations and, therefore, the potential for knowledge exploitation. Conceptualization is only useful for EDA if users' concepts cover all images and describe them exhaustively.

\subsection{Study Design}

\subsubsection{Data}
\label{sec:data}

\begin{figure}[t]
  \centering
  \begin{subfigure}[t]{0.32\columnwidth}
    \centering
    \includegraphics[width=\textwidth]{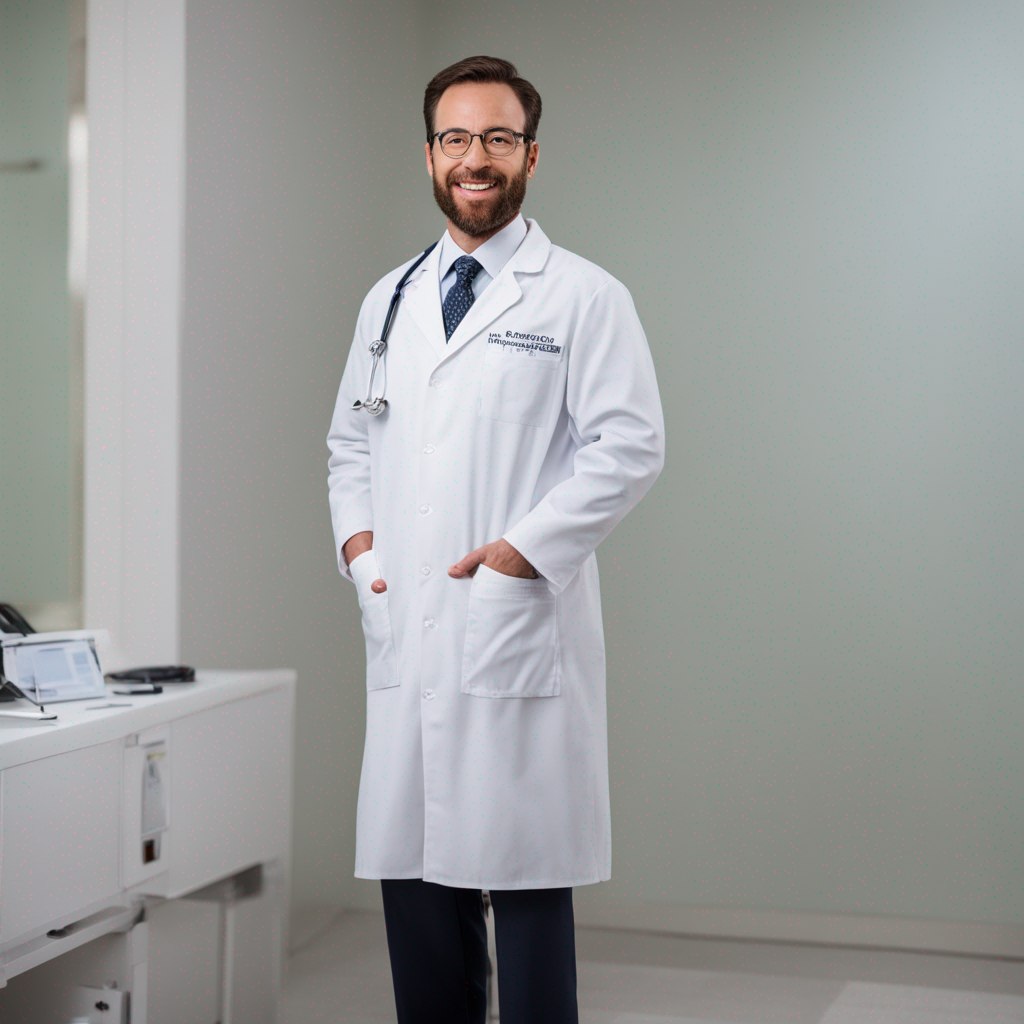}
    \caption{physician}
  \end{subfigure}
  \begin{subfigure}[t]{0.32\columnwidth}
    \centering
    \includegraphics[width=\textwidth]{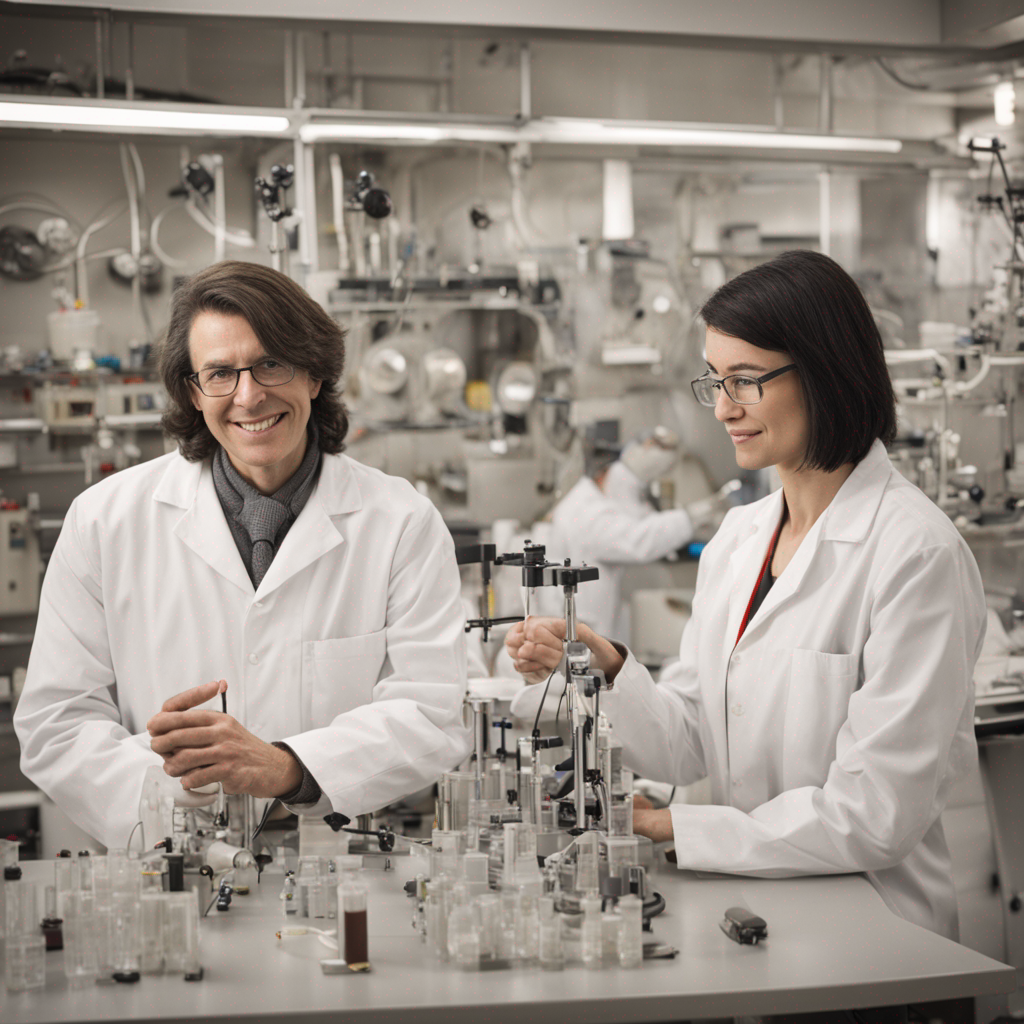}
    \caption{physical scientist}
  \end{subfigure}
  \begin{subfigure}[t]{0.32\columnwidth}
    \centering
    \includegraphics[width=\textwidth]{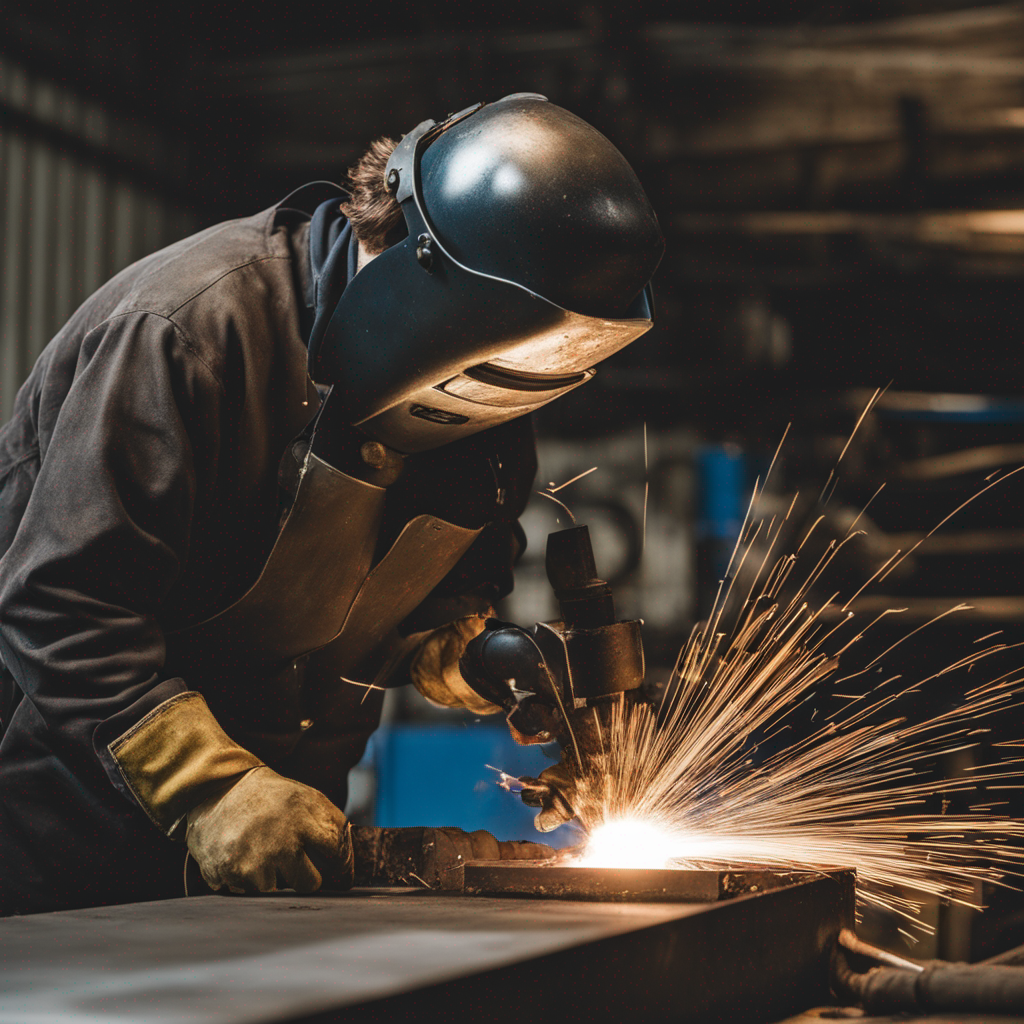}
    \caption{welding worker}
  \end{subfigure}
  \begin{subfigure}[t]{0.32\columnwidth}
    \centering
    \includegraphics[width=\textwidth]{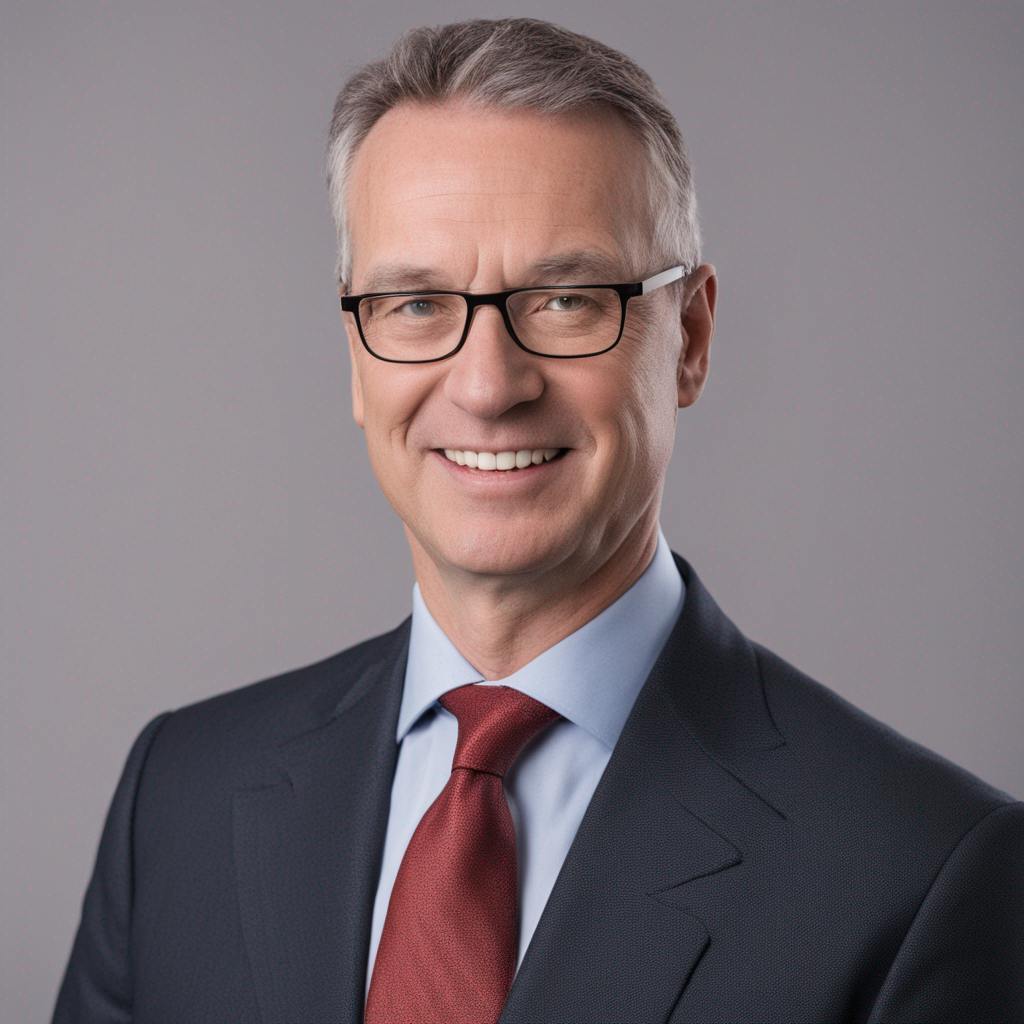}
    \caption{chief executive}
  \end{subfigure}
  \begin{subfigure}[t]{0.32\columnwidth}
    \centering
    \includegraphics[width=\textwidth]{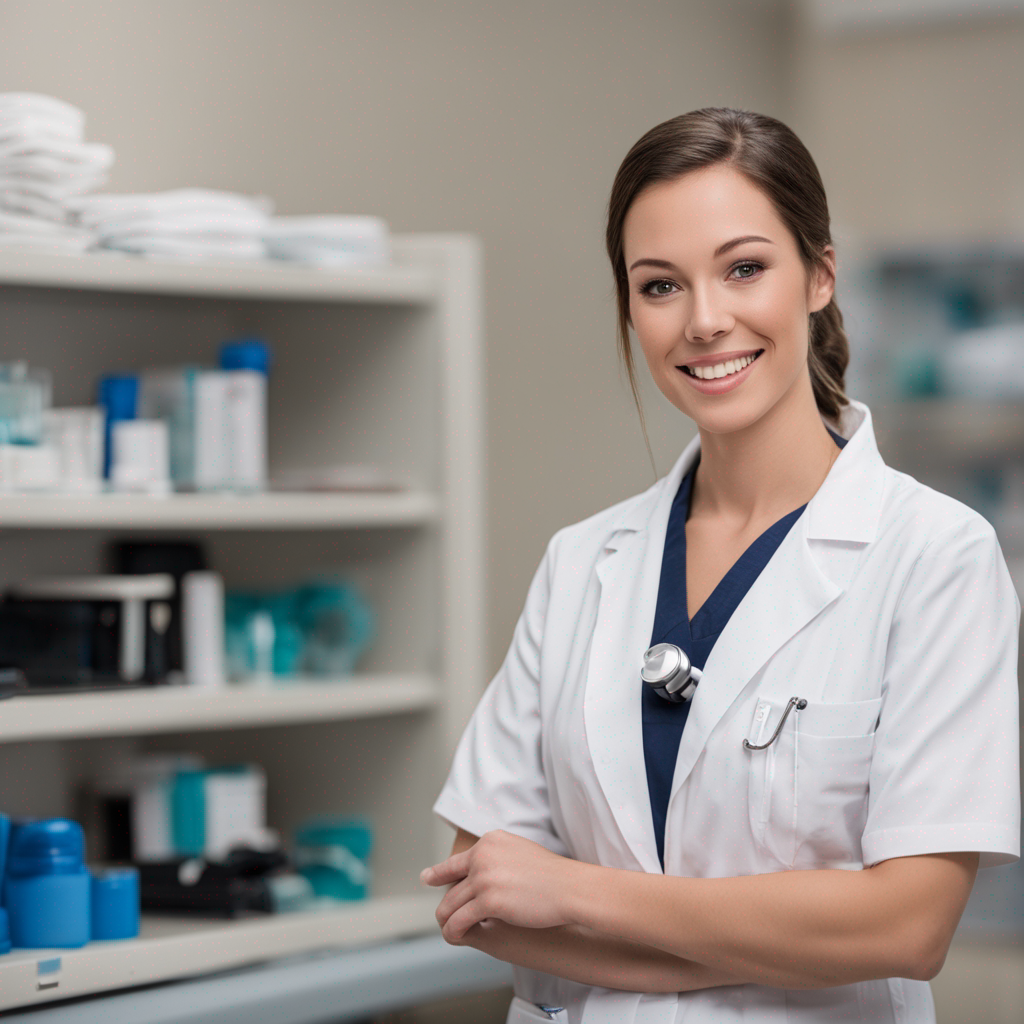}
    \caption{medical assistant}
  \end{subfigure}
  \begin{subfigure}[t]{0.32\columnwidth}
    \centering
    \includegraphics[width=\textwidth]{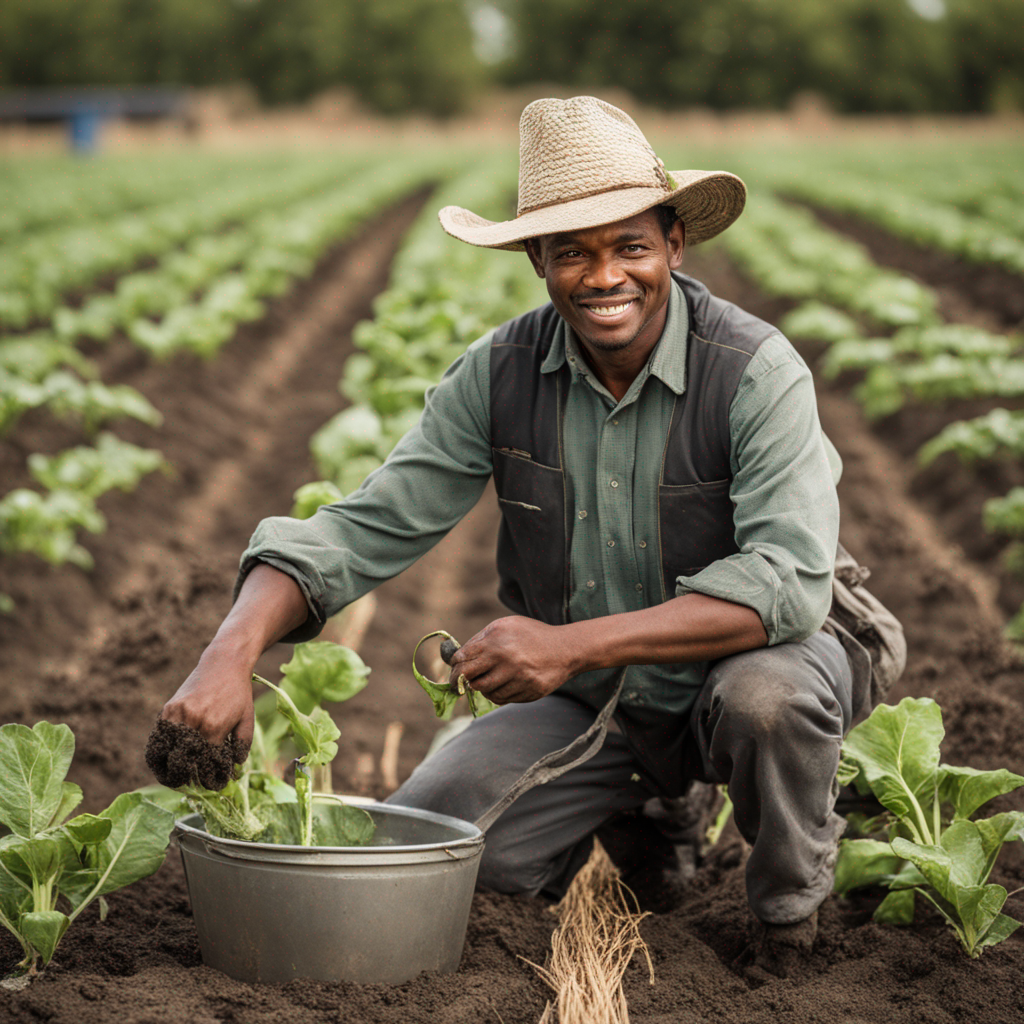}
    \caption{agricultural worker}
  \end{subfigure}
  \caption{Example images from the user study generated with Stable Diffusion XL, showing people from different occupations. Note that the potential biases present in the generated images were a deliberately chosen feature of the dataset.}
  \Description{Six example images, labeled (a) to (f), from the user study, showing people from different occupations. The first image (a) shows a physician with glasses and a beard, wearing a white lab coat, standing in a mostly empty room. The second image (b) shows two people working in a lab. Both wear white lab coats, have long hair, and wear glasses. On the table in front of them and in the background, various lab equipment is scattered around. Image (c) shows a welder with their face covered with a welding mask. The welder is looking down at a workpiece as sparks fly. Image (d) shows a smiling man with gray hair and glasses, wearing a suit and a red tie. The background is gray. Image (d) shows a young woman in a lab coat standing in front of a shelf with medical supplies. She has long brown hair and is smiling. Image (f) depicts a black man in a field, smiling at the camera; behind him are rows of plants, and next to him is a bowl of harvested plants. He is smiling at the camera, wearing a tan hat and a green shirt.}
  \label{fig:stimuli}
\end{figure}

The image dataset was generated using Stable Diffusion XL~\cite{podell_sdxl_2023}. A sample of the dataset is shown in Figure~\ref{fig:stimuli}. Note that this data is clearly not \emph{large} unstructured data. We deliberately chose a small-scale dataset for two reasons. Firstly, as we needed to obtain a human-curated ground-truth mapping between data and externalized concepts via crowdsourcing, we had to limit the number of images. Secondly, we did not provide any support for users during the exploration process through knowledge exploitation; thus, the limited dataset helped prevent users from being overwhelmed. The study dataset comprised the 100 most common U.S.~occupations (Bureau of Labor Statistics~\cite{bureau_of_labor_statistics_labor_2023}), selected at the finest granularity and ranked by \emph{total employed}. Occupation names were standardized into singular form (e.g., ``chief executives'' → ``chief executive'') and, where needed, slightly adapted to avoid ambiguity in the text-to-image prompts. Each prompt followed the format: ``A professional photograph of a [occupation]'' resulting in one image per occupation (see \href{https://osf.io/xhdmv/files/j5adt}{our OSF repository} for the full list).

\begin{figure*}[t]
    \centering
    \includegraphics[width=0.6\textwidth]{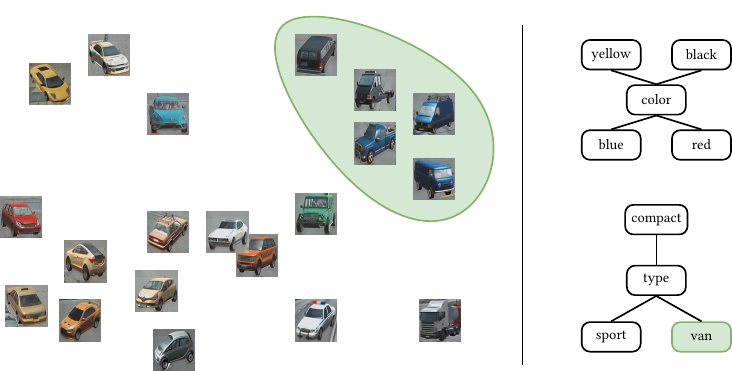}
    \caption{Illustration of how we determined whether concepts were spatially separated on the table, exemplified with the cars dataset~\cite{vedaldi_simulating_2020} used for participant training, and the concept ``van''. In this example, the vans form a well-separated cluster on the table, potentially resulting in a high rGON score.}
    \Description{The figure is split into two parts: on the left, scattered car images are displayed, representing different types of cars. A group of five blue van-like vehicles in the upper right area is highlighted with an ellipse. On the right, a concept map with two facets is shown. The facet labeled ``color'' connects to four concepts: yellow, black, blue, and red. The second facet labeled ``type'' connects to three nodes: compact, sport, and van. The ``van'' node is highlighted with a green outline and fill, matching the highlighted car group on the left.}
    \label{fig:cars}
\end{figure*}

To help users familiarize themselves with the task and interfaces, they first performed a training task with a smaller set of images showing 20 synthetic car renderings with clearly distinct visual attributes, such as color, orientation, or car type~\cite{vedaldi_simulating_2020}. This setup is illustrated in Figure~\ref{fig:cars}.

\subsubsection{Apparatus}
\label{sec:apparatus}

The exploration interface was a canvas displayed on a 55-inch multi-touch table with a 5K display, showing all images from the dataset (see Figure~\ref{fig:study_setup}). The table was positioned and tilted so that participants could comfortably interact with it while standing.

The images were initially organized as a single stack of 100 images in the center of the table, with only the uppermost image being visible. This allowed us to track how many images a participant had seen at a given time step. The image order was randomized once after generation, and every participant was then presented with the \textit{same} order to enable comparing exploration processes over time. Images were shown with a resolution of $100 \times 100$ pixels.

To create the concept map, users were provided with a separate device, a 13-inch laptop with a trackpad and an external mouse. Users started with an empty canvas and could create shapes with text labels (concepts) and connect them with undirected edges to represent relationships between the concepts. Participants could create, delete, move, and rename concepts, create edges, connect concepts with edges, and label edges.

\subsubsection{Procedure}
\label{sec:procedure}

The study started with an introduction and the signature of a consent form, followed by a survey on demographics and prior experience with visualization tools and concept maps. Afterward, the participants were introduced to the interfaces via a tutorial task. Participants were encouraged to employ \emph{thinking aloud}, verbalizing their thoughts and actions as they worked.

For the main task, users were provided with the 100 images described in Section~\ref{sec:data}. Users were informed of the expected duration (45 minutes) and reminded of the remaining time after 40 minutes. After the task, we conducted a semi-structured interview with questions concerning the participant's experience, difficulties, and thoughts about the task. Additionally, the participants were asked to explain the concepts they discovered, their concept map, and the spatial organization on the touch table.

During the task, the researchers took notes, and the entire task was recorded in audio and video for qualitative retrospective analysis. All interactions across both interfaces were logged comprehensively, allowing the entire session to be replayed later.

\subsubsection{Participants}

We recruited 20 participants from the staff and student body of a local university from different IT-related fields (computer science, data science, and digital health care). Twelve participants were between 24 and 35 years old, seven were between 18 and 24, and one was over 35. Nine participants identified as men, nine as women, and two as non-binary. Twelve participants had an academic degree (bachelor's or higher), and eight held a high school degree. The majority of the participants (13) had used concept maps or mind maps before. Participants were compensated with 20 Euros for an average total time spent of 77 minutes (min 52, max 91).

\subsection{Data Curation}
\label{sec:dataCuration}

It was necessary to manually curate the artifacts created by the users in the two conceptualization interfaces for multiple reasons: (1) to establish a ground-truth assignment between the images and the concepts expressed in the concept maps for assessing the reliability of a machine-assisted ``search for structure'' (\textbf{RQ2}), (2) to quantify the extent of spatial organization of the images on the table with respect to the discovered concepts (part of \textbf{RQ1}), and (3) to assess whether the discovered concepts describe the images well enough to support queries and filtering (\textbf{RQ3}). 

Two co-authors, acting as independent coders, therefore, first established a cleaned \textit{superset} of concepts from the final concept maps. 
In their concept maps, participants externalized 698 concepts in total through node labels. 
The two coders individually analyzed the text labels of all nodes in all users' concept maps following methods from qualitative and formal concept analysis~\cite{al-diban_comparison_2011}. Non-English node labels were first translated into English and then coded in two iterations. In the first iteration, nodes representing more than one concept were split (e.g., a node labeled ``charts, books, boxes'' was split into three distinct concepts), unspecific or incomprehensible labels (e.g., ``other'' or ``full body pose'') were ignored, and labels with equivalent semantics were merged into a single concept -- for example, ``a man'', ``men'', ``male'', and ``m'' were merged into ``male''. In the second iteration, each concept was then assigned to a single facet, or, if no suitable facet was present in the data, a new facet was introduced by the coders to fill obvious gaps or make implicitly defined facets explicit. In both steps, disagreements in concept codes were discussed until the coders reached a consensus. At the end of the coding process, a total of 343 concepts and 70 facets (seven of which were defined by the coders) were identified.

In the next step, we used crowd workers to establish a ground truth assignment between the 343 superset concepts and the 100 images (resulting in 34,300 image-concept assignments).  
We recruited a total of 146 crowd workers from Amazon Mechanical Turk (MTurk), who were compensated at an average hourly rate of 14 USD. In short human intelligence tasks (HITs), we asked workers whether a batch (10 images) contained a concept in the context of its associated facet (``Does this image contain [concept]?''). The answer options for each image were \textit{Yes}, \textit{No}, and \textit{I don't know}. The last option was provided in case the concept is not clearly visible in the image or when the question is ambiguous for the given image, for instance, for more abstract concepts like ``chemistry''. Each image-concept combination was evaluated by at least three workers. 
With the neutral response option, however, the vote could remain ambiguous. In that case, an expert vote was provided by one of the co-authors, who was unfamiliar with the superset of concepts. This tie-breaking was necessary for 497 concept-image pairs ($\sim1.4\%$). If the expert vote was also inconclusive (which occurred in three cases), the concept was considered absent in the image.

\subsection{Analysis}
\label{sec:analysis}

For the exploratory analysis of the study data, we followed a concurrent mixed-methods design emphasizing the quantitative part with an embedded qualitative component~\cite{leech_typology_2009}. Quantitative analysis of the interaction logs, the touch table layouts, and concept maps formed the basis for answering our research questions, while qualitative analysis of the post-task interviews, interaction replays, and video recordings was used to contextualize, explain, and interpret the quantitative findings. For the coding of the qualitative part, we utilized reflexive thematic analysis~\cite{braun_reflecting_2019}. The guiding questions of the analysis were developed after the two coders initially familiarized themselves with the data. This was followed by independent coding of the transcribed interview data. Coders then discussed the codes and unified them into a codebook, including only codes present in more than one interview. Initial themes were then defined based on the unified code book, and in a final iteration, the relevance of each theme was determined by both coders. In some cases, the researchers' notes and video recordings were used to clarify or validate the participants' interview statements during the analysis. 
All data and analysis scripts are \href{https://osf.io/xhdmv}{available on OSF}.

\subsubsection{RQ1: Knowledge Organization Strategies}
\label{sec:analysis-rq1}

To answer \textbf{RQ1} on the knowledge organization strategies applied by the users, we performed the following analyses: 

\paragraph{Exploration Sequence} We analyzed the temporal sequence of image viewing from the touch table interaction logs and conceptualization in the concept map from the concept map interaction logs. If users followed a strict top-down classification, we expect they would start by creating a conceptualization at the beginning of the exploration process. Conversely, when following a strict bottom-up classification approach, we expect users to first view all images before starting the conceptualization. Our quantitative analysis was supported by interview themes addressing the temporal interplay between exploration and conceptualization. 

\paragraph{Concept Map Organization}
To understand which structure users applied in their concept maps, we classified the graph topology from the final concept map logs into four different structural types based on the work by Richmond et al.~\cite{richmond_set_2014}: \emph{network}, \emph{hub-and-spoke}, \emph{linear/chain}, \emph{tree} (see Table~\ref{tab:structure-heatmap}). Isolated nodes were excluded. Because the concept map edges were undirected, we inferred directionality based on their categorization into concepts and facets. For each participant, the concept map was decomposed into its main structural components as follows: based on the inferred directionality, we identified all concepts with only outgoing links and subsequently assigned all remaining concepts to the aforementioned structural types based on their local connectivity patterns (e.g., multiple incoming links refer to a \emph{network} structure, one incoming and multiple outgoing links represent a \emph{tree} structure). For each participant, we computed a structure score for each type, shown in Table~\ref{tab:structure-heatmap}, as the proportion of concepts assigned to that type relative to the total number of concepts, excluding stand-alone concepts. If users tried to create strict taxonomies, we would observe tree structures. Faceted classification would yield hub-and-spoke structures. A pure tagging approach would result in isolated nodes with no connections. This analysis was supported by interview themes that addressed the construction process of the concept maps. 

\paragraph{Spatial Organization}
To understand whether and according to which semantics users spatially structured images on the tabletop interface, we quantified the number of concepts whose related images were spatially separated from others (i.e., clustered) in the final spatial arrangement on the touch table using the touch table logs. For each user $u_i$, we assessed whether the set of images $I(c_k)$ associated with a concept $c_k$ is closer to each other than to the images not associated with the concept $I \setminus I(c_k)$, as illustrated in Figure~\ref{fig:cars}. There are multiple spatial separation metrics for approximating perceived visual separation~\cite{tatu_combining_2009, sips_selecting_2009}. The GON score~\cite{aupetit_sepme_2016} has been shown to closely match human judgments of spatial separation. Since the GON score assumes classes to be balanced, we derived the residual GON score ($rGON$) by subtracting the fraction of data items known to be associated with $c_k$:  
\begin{equation}
    rGON(u_i, c_k) = GON(u_i, c_k) - \frac{|I(c_k)|}{|I|}. 
    \label{eq:rgon}
\end{equation}
We computed all $343 \times 20 = 6,860$ rGON scores for all 343 concepts in the concept superset from the final table arrangements of the 20 users. The median score in our experiment is .0085, and the maximum score is .99. By qualitatively comparing the distribution of rGON scores to the final table arrangements and the users' reports in the post-experiment interviews, we found that only extreme outliers (i.e. values more than three interquartile ranges ($IQR$) above the third quartile ($Q3$)) of the rGON score represent concepts that have been intentionally expressed through spatial separation. In our experiment, the extreme outlier threshold was $t_s = Q3 + 3 \cdot IQR = 0.42$. Thus, we define a spatial categorization corresponding to a concept $c_k$ on user $u_i$'s table arrangement if 
\begin{equation}
    rGON(u_i, c_k) \geq t_s.
    \label{eq:rGON_threshold}
\end{equation}

\paragraph{Agreement}
To see whether users derived a classification from an initial categorization, as suggested by Rosch~\cite{rosch_basic_1976}, we quantified how many concepts in a user's concept map are spatially separated on the tabletop, thereby computing the agreement between the two conceptualizations. If users mapped their observed categories to the concept map, there should be at least one concept in their concept map that is also spatially distinct on the tabletop. 

\subsubsection{RQ2: Machine-Assisted ``Search for Structure''}
\label{sec:analysis-rq2}

We used a pre-trained CLIP~\cite{radford_learning_2021} model to test zero-shot assignment and automatic semantic categorization capabilities of a well-known, representative state-of-the-art multimodal model in a post-hoc fashion. Intuitively, a pre-trained model will provide solid performance for well-known concepts but is expected to perform worse for underrepresented ones~\cite{radford_learning_2021}. Consequently, we performed our analyses both for the entire concept superset and for the subset containing only the most popular concepts. For this distinction, we defined \emph{popular concepts} as those that appeared in at least 25\% of the concept maps (5/20). This is the case for 24 out of the 343 concepts in the superset. 

\paragraph{Zero-Shot Assignment Reliability}
For each concept $c_k$ from the concept superset, we compared the similarity measures computed in CLIP space (as defined in Equation~\ref{eq:clip-sim}) between the set of images $I(c_k)$ assigned to the concept $c_k$ and the set of images $I \setminus I(c_k)$ not assigned to $c_k$ with a Mann-Whitney U test.  
We computed the effect sizes through the standardized z-scores $r=\frac{Z}{\sqrt{|D|}}$ for each concept, where $D$ is the set of all images. For each user, we report the number of concepts that yield at least a moderate effect ($|r|\geq0.3$).

\paragraph{Automatic Semantic Categorization}
For each concept $c_k$, we computed the $rGON(p,c_k)$ score (Equation~\ref{eq:rgon}) that quantifies the visual separation of $c_k$ in the UMAP projection $p$ of all images' CLIP embedding vectors. We use our empirically selected threshold $t_s$ (Equation~\ref{eq:rGON_threshold}) to decide whether $c_k$ is sufficiently spatially separated in the projection space so that it can be considered to represent a detectable semantic category. 

\subsubsection{RQ3: Exploratory Analysis}
\label{sec:analysis-rq3}

To answer \textbf{RQ3} concerning the potential for exploiting the users' conceptualizations for actual exploratory data analysis (Figure~\ref{fig:teaser}, left), we performed the following analyses: 

\paragraph{Exhaustiveness}
Concepts describe the data well if each image is associated with at least one externalized concept. To assess how well the data is covered by the user's concept map, we compute the fraction of the 100 images that are not assigned to any concept in the user's concept map, based on our ground-truth mapping. A value of 1.0 means that each data item is described by at least one concept; 0.0 would mean that not a single image can be assigned to a concept. If exhaustiveness is $< 1.0$, exploratory analysis will be unreliable, since some parts of the data lack usable descriptions and may therefore not be included in summary statistics. 

\paragraph{Informativeness}
More importantly, concepts describe the data well if each image can be described by a unique combination of concepts. We define informativeness as the ratio of images described by a \textit{unique} combination of concepts ($[0.0;1.0]$). An image is considered uniquely described if there exists a subset of concepts in the user's concept map that is associated with no other image. A higher value allows more precise querying. In theory, at least seven concepts are required to uniquely identify all 100 images, as seven distinct concepts can encode $2^7=128$ unique combinations.

\section{Results}
\label{sec:results}

\subsection{RQ1: Knowledge Organization Strategies}
\label{sec:res-rq1}

\subsubsection{Exploration Sequence}
\label{sec:res-sequence}

\begin{figure*}[ht]
    \centering
    \includegraphics[width=\textwidth]{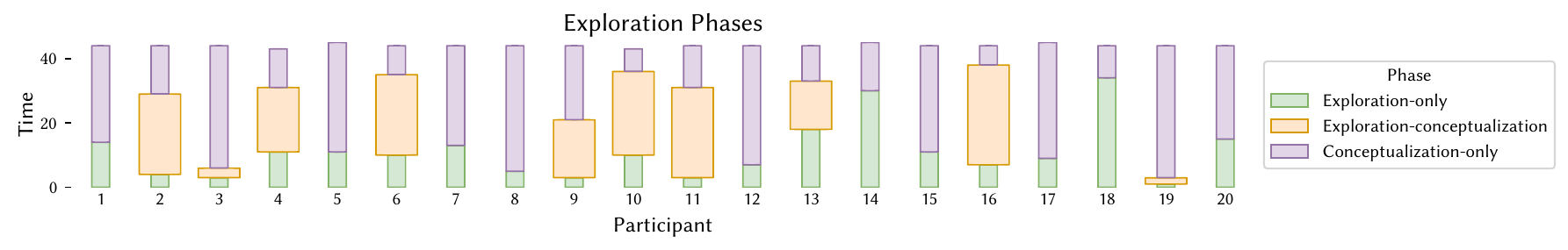}
    \caption{Time spent in each phase per participant in minutes. During \textit{exploration-only}, images were viewed but no concepts are added; in \textit{exploration-conceptualization}, concepts are added while new images are uncovered, and in \textit{conceptualization-only}, all images have already been uncovered, but new concepts are still being added.}
    \Description{A chart consisting of 20 line charts stacked on the y-axis. Each line represents one participant. The individual charts are segmented along the x-axis. The x-axis represents time, ranging from 0 to 45 minutes. The three different segments represent the phases the users went through during the study. Nine of the charts (participants 1, 5, 7, 8, 12, 14, 15, 17, 18, and 20) lack the middle segment that indicates the exploration-conceptualization phase. All 20 participants roughly used the same amount of time, 45 minutes, as indicated by the cumulative lengths of the individual line charts.}
    \label{fig:externalization-phases}
\end{figure*}

None of the users created concepts without first looking at at least some images. The lowest number of images viewed before initiating conceptualization was eight. In the interview, two users explicitly mentioned prioritizing getting an overview over immediate interaction. The largest group of users (12/20) followed -- what was described in the interview as an ``\emph{all-at-once}'' strategy: within an average time period of only ten minutes, they had uncovered all images, but afterward spent an average of over 30 minutes carefully creating the concept map -- twice as long as the remaining eight users. A smaller number of four users described their strategy as ``\emph{batched}'' by uncovering batches of, for instance, ten images and creating corresponding concepts before repeating the operation with the next batch. The four remaining users described their strategy as ``\emph{gradually}'' uncovering images one after the other while simultaneously creating concepts in the concept map whenever a novel concept was discovered. 
Figure~\ref{fig:externalization-phases} shows how the 20 users transitioned between three phases: (1) exploring images only, (2) exploring images while conceptualizing, (3) conceptualizing only without any interaction with the images. 

In summary, all users had a look at -- at least a couple of -- the images before starting the conceptualization. This strongly indicates that concepts were defined in a \emph{bottom-up classification} manner by the users. About half of the users looked at \emph{all} images before even attempting to build a concept map. 

\subsubsection{Concept Map Organization}
\label{seq:res-conceptmap-organization}

\begin{figure}[ht]
  \centering
    \includegraphics[width=\columnwidth]{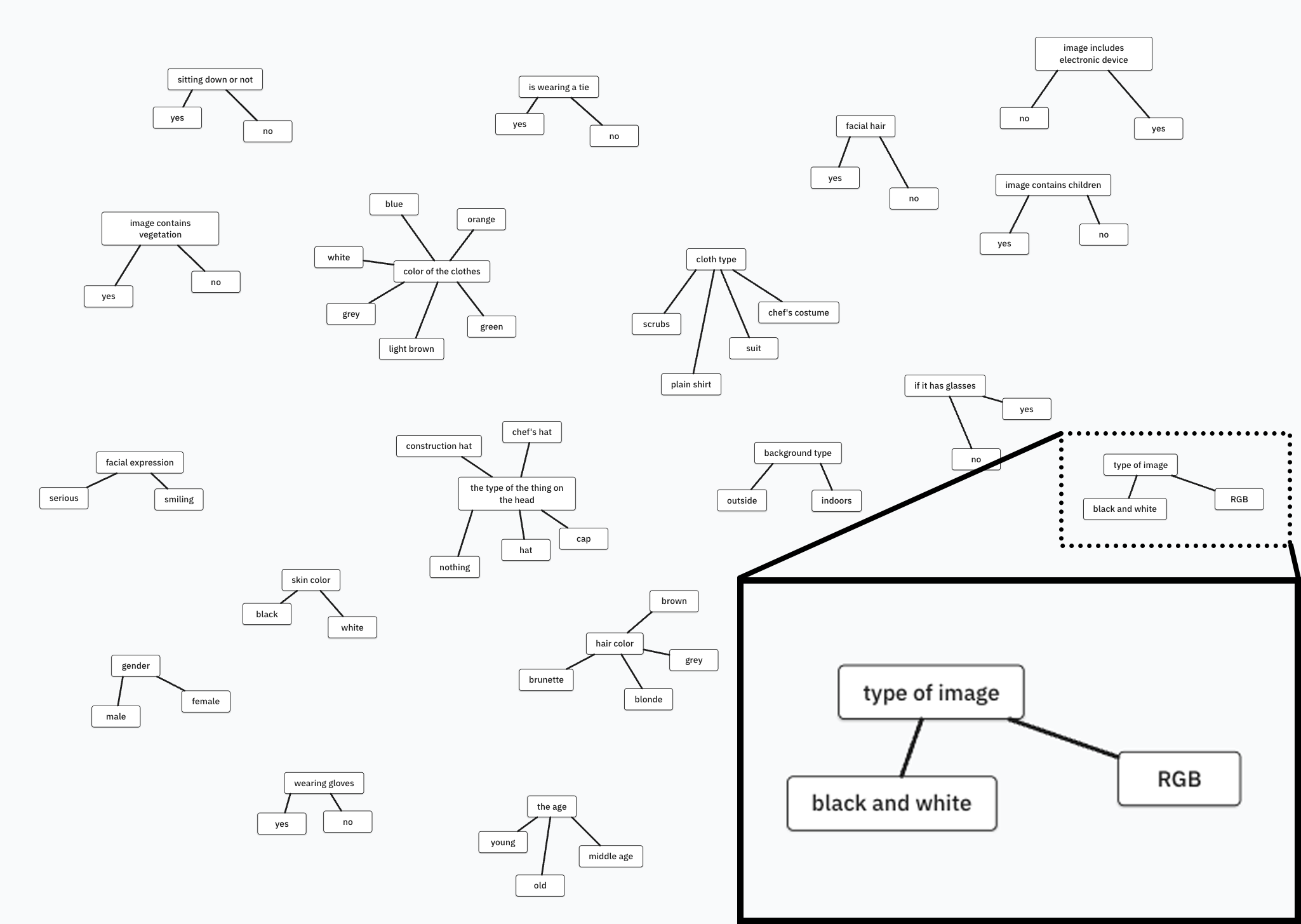}
  \caption{Concept map relying solely on the most commonly observed hub-and-spoke structure (participant 17).}
  \label{fig:conceptMaps}
  \Description{A collection of node-link diagrams, each with a central node linked to additional nodes in a star-like hub-and-spoke structure. The nodes represent concepts discovered by the user. One example is highlighted via an inset: the hub ``type of image'' with the child nodes ``black and white'' and ``RGB'' (i.e., color image).}
\end{figure}

All users created some nodes in their concept maps. Overall, the concept maps contained between 9 and 94 concepts (median 28). 14 out of our 20 users also created nodes that were classified as facets by our coders. The median number of nodes classified as facets per user was six. 
Most participants created concept maps with multiple structural types, but with one structure clearly dominant. 14 users predominantly created multiple unconnected \emph{hub-and-spoke} structures, in which the central hub node typically represented a facet such as \emph{background} or \emph{facial expression} (Table~\ref{tab:structure-heatmap} and Figure~\ref{fig:conceptMaps}). Only two participants created notable \emph{network} structures, which typically indicate a deeper understanding in concept mapping~\cite{kinchin_how_2000}. 
In 12/20 concept maps, there was at least one isolated node. The median fraction of isolated nodes was low (5.9\%); i.e., most nodes were connected by edges. However, the semantics of edges were generally rather inconsistent in our study. Also, the spatial layout of most concept maps did not imply any apparent semantics. 4/20 users reported representing relations by proximity instead of edges. These concept maps contained more than 50\% isolated nodes. 

In the interview, 4/20 users explicitly reported that finding appropriate labels for the concept map nodes was challenging. 
The process of creating the concept map was described as rather \emph{top-down} by a majority of users (12/20) in our interview. 
That means, most users would describe their workflow as first naming a facet (e.g., \emph{``hair color''}) and then later populating it with specific concepts (e.g., \emph{``brown''}). 
Users who described their process as a top-down approach also had significantly more concepts in their concept maps than those who described their approach as completely \emph{unstructured} (median number: 45 vs. 20; $U=80.5; p=.010$). 

In summary, the majority of the observed concept maps (16/20) were too fragmented and lacked sufficient depth to represent a strictly hierarchical taxonomy. At the same time, they were too structured to be considered a flat tagging approach. The majority of users (14/20) therefore seemed to construct a \emph{faceted classification} scheme, and most of these users (12/14) created their concept maps in a \emph{top-down mapping} approach -- from general facets to specific concepts.

% Narrow table optimized for double column layout

\begin{table}[ht]
    \centering
    \small
    \begin{tabularx}{\columnwidth}{p{0.1cm}p{0.1cm}>{\centering\arraybackslash}X>{\centering\arraybackslash}X>{\centering\arraybackslash}X>{\centering\arraybackslash}X}
        \multirow{24}{*}{\rotatebox{90}{\textbf{Participant}}} & & \includegraphics[width=1cm]{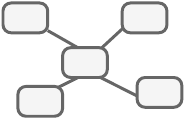} & \includegraphics[width=1cm]{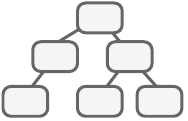} & \includegraphics[width=1cm]{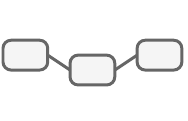} & \includegraphics[width=1cm]{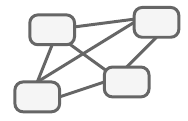} \\
        & & \textbf{Hub \& Spoke} & \textbf{Tree} & \textbf{Linear} & \textbf{Network} \\
        & 1  & \cellcolor[RGB]{28,105,137} \textcolor{white}{0.68} & \cellcolor[RGB]{99,172,144} 0.24 & \cellcolor[RGB]{143,194,145} 0.08 & \cellcolor[RGB]{165, 205, 144} 0.00 \\
        & 2  & \cellcolor[RGB]{44,53,115} \textcolor{white}{0.98} & \cellcolor[RGB]{159,202,145} 0.02 & \cellcolor[RGB]{165, 205, 144} 0.00 & \cellcolor[RGB]{165, 205, 144} 0.00 \\
        & 3  & \cellcolor[RGB]{39,69,124} \textcolor{white}{0.89} & \cellcolor[RGB]{165, 205, 144} 0.00 & \cellcolor[RGB]{165, 205, 144} 0.00 & \cellcolor[RGB]{134,190,145} 0.11 \\
        & 4  & \cellcolor[HTML]{1e5c86} \textcolor{white}{0.76} & \cellcolor[HTML]{80bb91} 0.13 & \cellcolor[HTML]{86be91} 0.11 & \cellcolor[HTML]{a5cd90} 0.00 \\
        & 5  & \cellcolor[HTML]{225282} \textcolor{white}{0.82} & \cellcolor[HTML]{89bf91} 0.10 & \cellcolor[HTML]{94c591} 0.06 & \cellcolor[HTML]{9fca91} 0.02 \\
        & 6  & \cellcolor[HTML]{27457c} \textcolor{white}{0.89} & \cellcolor[HTML]{9ac791} 0.04 & \cellcolor[HTML]{92c491} 0.07 & \cellcolor[HTML]{a5cd90} 0.00 \\
        & 8  & \cellcolor[HTML]{1e5c86} \textcolor{white}{0.76} & \cellcolor[HTML]{63ac90} 0.24 & \cellcolor[HTML]{a5cd90} 0.00 & \cellcolor[HTML]{a5cd90} 0.00 \\
        & 9  & \cellcolor[HTML]{26497e} \textcolor{white}{0.87} & \cellcolor[HTML]{9ac791} 0.04 & \cellcolor[HTML]{a5cd90} 0.00 & \cellcolor[HTML]{8bc191} 0.09 \\
        & 10 & \cellcolor[HTML]{33858d} \textcolor{white}{0.50} & \cellcolor[HTML]{5fa990} 0.26 & \cellcolor[HTML]{84bd91} 0.12 & \cellcolor[HTML]{84bd91} 0.12 \\
        & 11 & \cellcolor[HTML]{2a3d78} \textcolor{white}{0.93} & \cellcolor[HTML]{92c491} 0.07 & \cellcolor[HTML]{a5cd90} 0.00 & \cellcolor[HTML]{a5cd90} 0.00 \\
        & 12 & \cellcolor[HTML]{a5cd90} 0.00 & \cellcolor[HTML]{a5cd90} 0.00 & \cellcolor[HTML]{33858d} \textcolor{white}{0.50} & \cellcolor[HTML]{33858d} \textcolor{white}{0.50} \\
        & 13 & \cellcolor[HTML]{1c6689} \textcolor{white}{0.70} & \cellcolor[HTML]{66ae90} 0.23 & \cellcolor[HTML]{92c491} 0.07 & \cellcolor[HTML]{a5cd90} 0.00 \\
        & 14 & \cellcolor[HTML]{1c6989} \textcolor{white}{0.68} & \cellcolor[HTML]{78b891} 0.16 & \cellcolor[HTML]{78b891} 0.16 & \cellcolor[HTML]{a5cd90} 0.00 \\
        & 15 & \cellcolor[HTML]{4e9d90} 0.34 & \cellcolor[HTML]{70b390} 0.19 & \cellcolor[HTML]{a5cd90} 0.00 & \cellcolor[HTML]{388a8e} 0.47 \\
        & 16 & \cellcolor[HTML]{287b8c} \textcolor{white}{0.57} & \cellcolor[HTML]{5fa990} 0.26 & \cellcolor[HTML]{75b690} 0.17 & \cellcolor[HTML]{a5cd90} 0.00 \\
        & 17 & \cellcolor[HTML]{2c3172} \textcolor{white}{1.00} & \cellcolor[HTML]{a5cd90} 0.00 & \cellcolor[HTML]{a5cd90} 0.00 & \cellcolor[HTML]{a5cd90} 0.00 \\
        & 18 & \cellcolor[HTML]{2b3a76} \textcolor{white}{0.95} & \cellcolor[HTML]{a5cd90} 0.00 & \cellcolor[HTML]{98c691} 0.05 & \cellcolor[HTML]{a5cd90} 0.00 \\
        & 19 & \cellcolor[HTML]{a5cd90} 0.00 & \cellcolor[HTML]{41918f} 0.42 & \cellcolor[HTML]{58a590} 0.29 & \cellcolor[HTML]{58a590} 0.29 \\
        & 20 & \cellcolor[HTML]{58a590} 0.29 & \cellcolor[HTML]{84bd91} 0.12 & \cellcolor[HTML]{2f818d} \textcolor{white}{0.53} & \cellcolor[HTML]{94c591} 0.06 \\
    \end{tabularx}
    \caption{Structure score for each concept map structure. The score for a given structure is computed as the proportion of concepts assigned to that structure relative to the total number of concepts, excluding stand-alone concepts.
    Participant 7 was excluded because no clear structure was evident.}
    \label{tab:structure-heatmap}
\end{table}

\subsubsection{Spatial organization}
\label{sec:res-spatial-organization}

\begin{figure}[ht]
  \centering
  \begin{subfigure}[b]{\columnwidth}
    \centering
    \includegraphics[width=\textwidth]{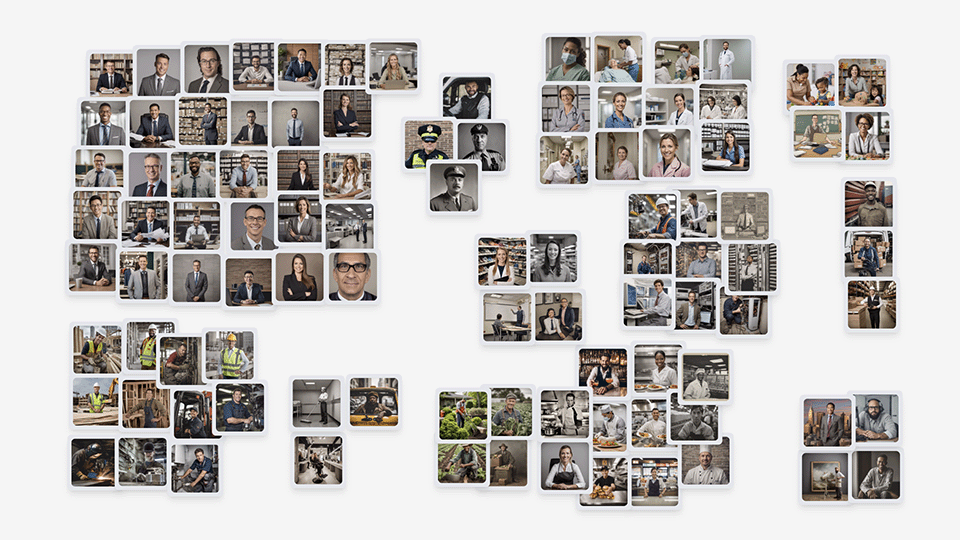}
    \caption{Semantic clusters in a grid-like layout (participant 3).
    }
    \label{fig:TT_clusters_grid}
  \end{subfigure}
  \begin{subfigure}[b]{\columnwidth}
    \centering
    \includegraphics[width=\textwidth]{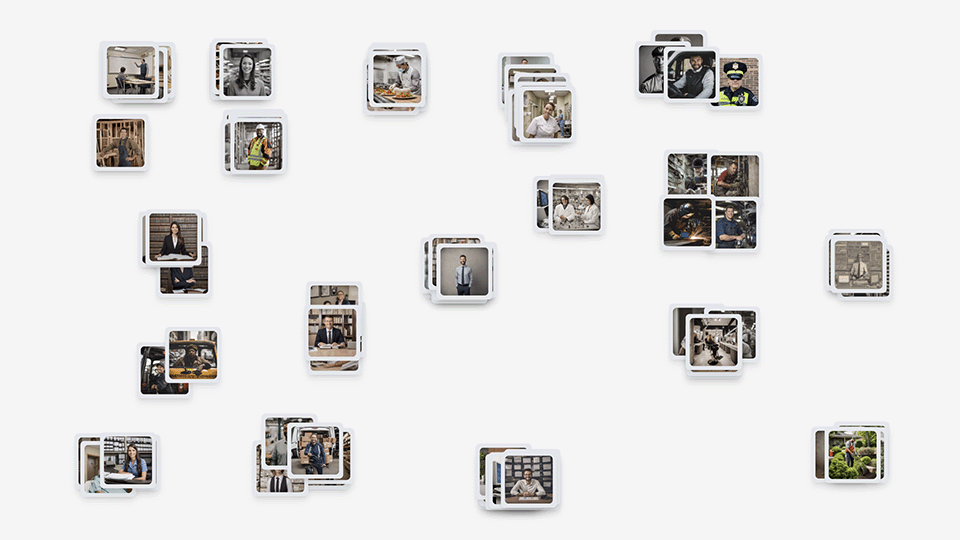}
    \caption{Semantic clusters in a pile layout (participant 14).
    }
    \label{fig:TT_clusters_pile}
    \end{subfigure}
    \begin{subfigure}[b]{\columnwidth}
    \centering
    \includegraphics[width=\textwidth]{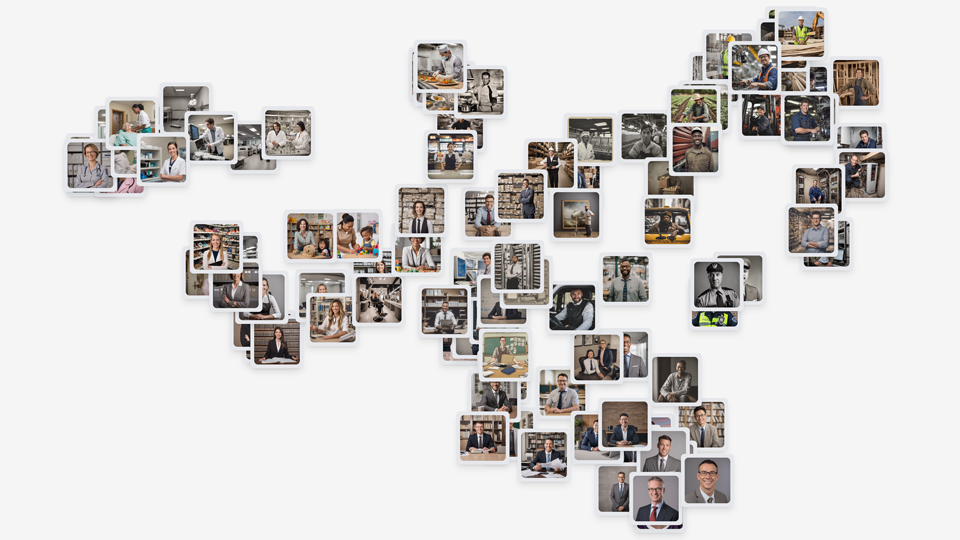}
    \caption{UMAP projection of the CLIP embeddings.
    }
    \label{fig:TT_umap}
  \end{subfigure}
  \caption{Two examples of spatial layouts from the touch table compared to the layout obtained via UMAP projection of CLIP embeddings.}
  \label{fig:touchTable}
  \Description{Two images showcasing layouts users constructed on the touch table and one showcasing an automatically generated layout obtained via UMAP. The first image (a) shows a grid-like layout in which all images are visible at the same time. There are 12 gridded clusters with whitespace between them. Within a cluster, the images are touching at their edges. The second image (b)  shows 19 stacks of photos with some whitespace between them. Only the topmost images are visible while the rest of the stack is occluded. This third image (c) is a UMAP projection of CLIP image embeddings for the 100 study images. Each point is displayed as a small thumbnail, so the plot shows the images themselves arranged by similarity in embedding space. The thumbnails form several clear clusters: office and meeting scenes are grouped in the center-left and lower-middle areas, construction and industrial worker images cluster in the upper-right, police/security-related images appear near the right-center, and additional mixed professional portraits and workplace scenes fill the center.}
\end{figure}

We found that 13/20 participants had at least one concept clearly reflected in their spatial organization of images. The median number of spatially separated concepts per user was five. The 13 users performing spatial organization had a median of 26 concepts in their concept maps, while the remaining seven users had a median of 44. This is not a significant difference ($U=38.0; p=.588$). 
According to the interview, the most important spatial organization aspect for many (11/20) participants was obtaining a clear overview of the data and seeing as many images as possible. Consequently, the majority of participants (15) self-reported organizing the images in a \emph{grid} or grid-like layout (see Figure~\ref{fig:TT_clusters_grid}). In contrast, five participants self-reported having the images organized in \emph{piles} (see Figure~\ref{fig:TT_clusters_pile}). 
Eight users explicitly mentioned in the final interview that creating and maintaining a spatial organization on the table was effortful.

In summary, a small majority of users performed \emph{categorization} in addition to populating their concept maps, which were only insignificantly smaller than the concept maps of users not performing any spatial organization. Spatial organization was therefore less in focus than the concept map. Indeed, subjective feedback suggests that maximizing visibility is more important than semantic categorization and that perceived effort is greater in the spatial organization of images.

\subsubsection{Agreement}
\label{sec:res-agreement}

Almost half of the users who performed semantic spatial organization (6/13) did not categorize the images on the table according to any concept present in their concept map.
That means only a few users (7/20) have potentially followed a strategy described by Rosch~\cite{rosch_basic_1976}, in which a classification is derived from an initial categorization. It seems that the categorization of images on the table and the concepts in the concept map are rather complementary.

\subsection{RQ2: Machine-Assisted ``Search for Structure''}
\label{sec:res-rq2}

\subsubsection{Concept and Category Popularity}
\label{sec:res-concept-popularity}

The ``search for structure'' was highly subjective: no single concept appeared in all concept maps, and around 60\% of all concepts were defined by only one participant. On average, each concept map contained 10 unique concepts. At the same time, certain concepts were relatively popular. In the concept maps, \emph{male} (15/20) and \emph{female} (14/20) were the most frequent, followed by objects (e.g., \emph{glasses}, \emph{food}) and salient visual attributes (e.g., \emph{black and white}, \emph{outdoor}), see also Table~\ref{tab:popularConcepts}. Less popular concepts often describe less prominent objects in single images (e.g., \emph{coffee}, \emph{fruit}, \emph{walkie-talkie}), poses (e.g., \emph{open hands}), or abstract concepts (e.g., \emph{industry}, \emph{art}, \emph{illness}). These results show that most concepts are idiosyncratic.

On the touch table, spatial organizations were similarly subjective, with 50/111 concepts (45\%) expressed by only one user. Most commonly, we observed a spatial separation by time period (\emph{1900s–1940s}, 6/13 who performed spatial organization). Despite explicit instructions to describe visual attributes irrespective of the professional context, concepts like \emph{healthcare}, \emph{farm}, or \emph{food retail} were also common, as can be seen in  Figure~\ref{fig:TT_clusters_grid} and Table~\ref{tab:popularConcepts}. 

\begin{table}[ht]
\centering
\small
   \begin{tabular}{l r r r}
   \toprule
    \textbf{Concept} & \textbf{~$X$~}  & \textbf{~$E$~} & \textbf{~$X$~$\cap$ $E$~} \\
    \midrule
    \textbf{Top-5 in concept map $X$} & & \\
    male                   & 15         & --    & --   \\
    female*\textsuperscript{\textdagger}                 & 14         & 4     & 4   \\
    headwear*\textsuperscript{\textdagger}               & 12         & 2     & 1  \\
    black and white*        & 11         & 5     & 3  \\
    glasses*                & 11         & --      & --  \\
    \midrule
    \textbf{Top-5 in spatial categorization $E$} & & \\
    1900s - 1940s                            &     1    & 6  & --  \\
    black and white*                          &     11   & 5 & 3\\
    healthcare*\textsuperscript{\textdagger}                               &     6  & 5  & 2 \\
    people of different genders              &     3 & 5 & 3\\
    pipes\textsuperscript{\textdagger}                                    &     1 & 5 & -- \\
    \bottomrule
   \end{tabular}
   \caption{The top-5 most popular concepts in number of participants (20 overall) expressing the concept explicitly in the concept map interface $X$~ (top) and implicitly in the exploration interface $E$~(bottom), as well as in both $X$~ and $E$. Concepts marked with * showed a moderate correlation in the zero-shot assignment experiment, those with \textdagger~ showed spatial separation in the embedding projection.}
   \label{tab:popularConcepts}
\end{table}

\subsubsection{Zero-Shot Assignment Reliability}
\label{sec:res-zero-shot-reliability}

We found a moderate correlation between CLIP similarity measures and ground-truth assignments for 4.7\% to 61.5\% of the concepts in the users' concept maps (median 20.3\%). 
In other words, for the majority of concepts in users' concept maps, CLIP would not reliably predict a higher similarity score for images associated with a given concept than for those not associated with it. For the 24 most popular concepts, we found a moderate correlation for 15 (see concepts marked with $*$ in Table~\ref{tab:popularConcepts}). As expected, this is a higher ratio than for the entire superset, but generally a lower reliability than expected and acceptable for a functional EluDA system.

\subsubsection{Automatic Semantic Categorization}
\label{sec:zero-shot-spatialization}

From the 343 concepts in the superset, we found 32 concepts that had an rGON score $\geq t_s$ in the UMAP projection (see Figure~\ref{fig:TT_umap}) of the CLIP image features -- and can therefore be considered to be visibly separate from the remaining concepts. Note that this is higher than the median number of manually created spatially distinct categories by participants. The most clearly distinct concept was \emph{doctor's office} ($rGON=.817$). Notably, most concepts with the highest rGON scores had a strong focus on professional contexts (e.g., \emph{chef's jacket, gastronomy, hospital, safety vest}). This partially correlates with popular categories observed on the manually organized tabletops (see Table~\ref{tab:popularConcepts}). 

\subsection{RQ3: Exploratory Analysis}
\label{sec:res-rq3}

\subsubsection{Exhaustiveness}
\label{sec:res-exhaustiveness}
Overall, exhaustiveness was high -- all but one participant had an exhaustiveness of 1.0, meaning every image was associated with at least one concept. The remaining participant’s concept map lacked concepts for two images, as most concepts were very specific. This means that all concept maps provided a solid foundation for subsequent exploratory analysis. 

\subsubsection{Informativeness}
\label{sec:res-informativeness}
Informativeness was also high, ranging from 0.88 to 1.0 (avg. 0.98), meaning practically all images could be described by a unique combination of concepts. Interestingly, we also found 11 concepts that did not appear in any image, including \emph{tattoos}, \emph{siren}, and \emph{amazon [the company]}, suggesting that some users also entered \emph{expected} concepts. 

In summary, this implies that users' conceptualizations would provide a fairly solid foundation for EDA in ideal conditions -- namely, when images are reliably assigned to concepts and users can obtain a representative overview of the entire dataset. 

\subsection{Discussion}
\label{sec:discussion}

We now revisit our research questions and discuss the results of our experiment with respect to our conceptual EluDA framework for the exploratory analysis of \emph{large} unstructured data: 

\paragraph{RQ1: Knowledge Organization Strategies}
\label{sec:dis-rq1}

The study reveals a strong tendency among users to perform \emph{bottom-up classification}, i.e., deriving a classification from the data while observing it, or -- more commonly -- even \emph{after} observing it. The most frequently observed structure in users' concept maps was small, isolated trees, indicating that many users applied a \emph{faceted classification} scheme. There is no evidence that users derived their classification from an initial categorization on the table. Instead, many participants expressed a desire to see as much of the data as possible to observe potential concepts. This approach to data exploration aligns with the open coding stage of grounded theory qualitative data analysis~\cite{lacey2009qualitative}, where solutions for human-AI collaboration are also emerging~\cite{jiang_supporting_2021, rietz_cody_2021}, making it a worthwhile field for finding inspiration on the distribution of agency between humans and AI.
Categorization was rare, matching findings from related work where it was shown that users refrain from spatialization when given the opportunity to explicitly externalize~\cite{waldner_linking_2021}. Furthermore, we observed that the concepts externalized were highly subjective, similar to the context of tagging, where it was found that only a few tags are used very commonly, while the majority are only used infrequently~\cite{halpin_complex_2007}. Occasionally, even concepts \textit{not visible} in the images were reported. Such concepts may originate from expectations that may reflect certain (social) biases. Both spatially arranging the data and finding appropriate labels for the concept map seem to require significant effort for users. This clearly calls for AI assistance to help users externalize and organize knowledge. 

\paragraph{RQ2: Machine-Assisted ``Search for Structure''}
\label{sec:dis-rq2}

In an ideal EluDA system, we expect the assignment step (see Figure~\ref{fig:teaser}) to work fully autonomously and reliably. In our experiment, zero-shot assignment using CLIP worked much less reliably than expected -- even for popular concepts, and even less for idiosyncratic ones. We assume that many of the user-defined concepts were either too abstract or too specific to be captured well by the CLIP image encoder. This shows that the assignment problem in EluDA cannot easily be solved by an established vision-language model. A more sophisticated strategy may be necessary.

Automatically discovered semantic categories were more meaningful. Interestingly, even though certain concepts appeared as semantic categories in the CLIP embedding projection (e.g., \emph{pipes}), the assignment of the given concept label to the images remained unreliable. This could indicate a semantic mismatch between users' descriptions and CLIP's learned textual representations. It could also indicate that users focus on fine-grained image details to discriminate them from others that CLIP struggles to capture. This is a known limitation of contrastively learned representations like CLIP, which use short image descriptions to learn the mapping between whole images and text~\cite{jing2024fineclip,monsefi_detailclip_2025}. Such descriptions tend to focus on global aspects of the images rather than fine details. Since there is little agreement between human and machine-based assignments when it comes to concepts concerning image details, a fully automatic ``search for structure'' approach does not seem promising for EluDA. Similarly, in the context of text annotation, recent work suggests that current LLMs should be used with caution, as their performance in unsupervised labeling can vary substantially~\cite{kristensen-mclachlan_are_2025}. The fact that most concepts were idiosyncratic indicates that conceptualization should remain under human control, calling for a conceptualization process driven by human agency and supported by AI.

\paragraph{RQ3: Exploratory Analysis}
\label{sec:dis-rq3}

Overall, the users' conceptualizations are detailed enough to support queries (every image was described by at least one concept) and more in-depth structured analysis (almost every image is described by a unique combination of concepts). Therefore, given a suitable assignment between concepts and data items, the conceptualizations fulfill the requirements necessary for applying traditional EDA methods~\cite{hartwig_exploratory_1979}. It seems that people strive to be exhaustive and informative, focusing on small details to differentiate content.
Note that, in our study, the dataset was small enough that users could look at the entire dataset within the study period. For a larger dataset, discovering discriminating concepts could be challenging without machine support. 

\section{EluDA Interface Requirements}
\label{sec:requirements}

%Based on the observations from our formative study, we derive interface requirements for EluDA systems: 
Our observations across \textbf{RQ1}-\textbf{RQ3} led us to four main challenges that future EluDA interfaces should address. Based on these, we derive the following interface requirements and discuss resulting opportunities in Section~\ref{sec:opportunities}.

\begin{enumerate}
    \item[\textbf{IR1}] \textbf{Maximize data visibility}: Our study has shown that all users aim to see (a lot of) the data before starting to conceptualize (Section~\ref{sec:res-sequence}) and most try to arrange the data to maximize their visibility (Section~\ref{sec:res-spatial-organization}). EluDA systems should, therefore, not require additional user effort to show as much of the data as possible. Conversely, they should show many representative samples from the entire dataset, with each sample covering both general trends and outliers. At the same time, they need to use the available display space wisely without overwhelming the user. 
    \item[\textbf{IR2}] \textbf{Minimize effort for the ``search for structure''}: Users reported that both, spatially arranging images and finding appropriate text labels for concepts, have been considered laborious (Sections~\ref{seq:res-conceptmap-organization} and \ref{sec:res-spatial-organization}). Even though the assignment has not been the subject of this study, manual labeling is generally considered a tedious task~\cite{wenyin_mialbum_2000}. EluDA systems should therefore generally aim to minimize the user effort for classification, categorization, and assignment. Clearly, though, our study suggests that full automation is not desirable due to a strong subjective component in the conceptualization step (Section~\ref{sec:res-concept-popularity}). It therefore requires a careful balance between human agency and automation.  
    \item[\textbf{IR3}] \textbf{Support conceptualization through faceted classification}: Even though we observed a variety of concept map organization approaches, faceted classification was the most frequently observed approach (Section~\ref{sec:res-concept-popularity}). Therefore, as a common ground, conceptualization interfaces should allow users to specify a set of facets, each containing multiple associated concepts. Further research is needed to determine whether a concept map interface like in our study, which allows users to build hub-and-spoke structures, is the best possible conceptualization interface for supporting this strategy. 
    In contrast to predefined ontologies or taxonomies, often used in knowledge-assisted systems~\cite{wagner_kavagait_2019}, our findings suggest that bottom-up faceted classification more closely reflects the users' reasoning. This should also be considered when designing AI-support techniques for the recommendation of potentially relevant concepts.
    \item[\textbf{IR4}] \textbf{Enable fast and reliable assignment}: According to our results (Section~\ref{sec:res-zero-shot-reliability}), fully automatic zero-shot assignment with established models such as CLIP is a fast but unreliable approach.   
    Observed concepts are often fine-grained and highly subjective. This asks for a collaborative human-AI approach in which concepts may be recommended by a respective model, and the user provides feedback on the relevance of the recommendations. In this collaboration, two essential aspects are that a) user effort is kept at a minimum (IR2); ideally, implicit feedback is leveraged to derive cues on a user's assessment of relevance, and b) the model should be adapted (and made sensitive) to the interest of the user over time via a fine-tuning mechanism.
    In the best case, the assignment step can blend into the conceptualization and EDA steps. What makes this particularly challenging is that users sometimes externalize ``expected'' concepts that may not be linked to any data item (Section~\ref{sec:res-informativeness}). Since user intervention should be possible in real time, zero-shot assignment needs to be fast enough to support interactive exploration.
\end{enumerate}

\section{Opportunities for Human-AI Collaboration}
\label{sec:opportunities}

In our formative study, no AI assistance was used. In our conceptual EluDA framework, the human retains agency, but machine support is required to process larger volumes of data. The study's findings enabled us to identify and describe opportunities for human-AI collaboration within the EluDA workflow. We present both human-AI collaboration approaches and traditional non-AI approaches that can be used to fulfill the requirements we identified in Section~\ref{sec:requirements}:

\paragraph{IR1: Maximize data visibility}
The requirement calls for effective decisions about \emph{what} data to show the user and \emph{how} to show it. It will be necessary to investigate sampling strategies that provide a representative subset of the data as a first overview to initiate the EluDA process. In our study, all users started exploring the data without providing any initial hints about their expectations. Consequently, sampling initially needs to be purely data-driven. Effective sampling could be enabled by (deep) active learning approaches~\cite{ren_survey_2021}, which could alternate between data- and model-driven selection strategies later in the analysis process. Seeing only a subset of the data may introduce various human biases during analysis, which should be taken into account when designing a sampling strategy~\cite{wall_warning_2017}. Alternatively (or in addition) to active learning, an initial clustering could be applied to identify representative items to start the process with (bootstrapping). 

The visualization of the representative data subset could reflect potential semantic categories by showing a two-dimensional projection of the data's embedding space, as in Figure~\ref{fig:TT_umap}. Projection variants that minimize both overlap and white space between data items~\cite{hilasaca_grid-based_2024}, are recommended. Projection views, thereby, do not need to remain static but could update the samples and the layout based on 1) concepts already reported by the user and 2) aspects in the data that are not well described by the user's current faceted classification (e.g., instances not well captured by already externalized concepts). 

\paragraph{IR2: Minimize effort for the ``search for structure''}
AI could support the user during their ``search for structure'' by 1) recommending categories by visualizing groups of semantically similar data items (see IR1), 2) suggesting concepts from a known concept vocabulary~\cite{hoseini2024survey}, as well as by 3) (semi-)automatically assigning data items to concepts and categories~\cite{pmlr-v133-desmond21a}.
Automatically extracting facets and concepts clearly introduces the risk of automation bias~\cite{wall_warning_2017}: reducing human intervention could lead users to focus only on popular concepts, resulting in less diverse conceptualizations and, ultimately, less diverse findings about the data. 
To counteract automation bias, the system may actively search for seemingly unimportant details in the data.
This is related to the concept of ``coverage'' in visualization~\cite{sarvghad_visualizing_2017}, which explicitly reveals to the user which dimensions in a structured dataset have already been covered by the user -- and which have not. Adapting this concept to unstructured data is not so straightforward, because the ``dimensions'' are initially unknown and only gradually discovered during the ``search for structure''.  

\paragraph{IR3: Support conceptualization through faceted classification}
Users seem to prefer conceptualizing bottom-up (i.e., deriving concepts from data observations), while mapping facets and concepts is mostly done in a top-down process (i.e., from general facets to specific concepts). When given a label for a facet, the EluDA system could suggest concepts semantically related to it. This could be accomplished by a conventional ontology or an LLM without even necessarily checking against the data. The same mechanism could be employed to complement an existing list of concepts provided by the user, either solely based on semantics or by letting a multimodal model check which facets and concepts could complement the description of existing data items. 

Following the two types of knowledge externalization as defined in the KAVA framework~\cite{federico_role_2017}, faceted classification is an explicit form of knowledge externalization that warrants a dedicated visual interface to make the conceptualization accessible and understandable to users. While we used a graph-like interface in our study to observe the variety of knowledge organization strategies, faceted classification could be represented by simpler structures, such as hierarchical lists. Especially if the EluDA system makes extensive suggestions, graph-like representations can easily become overwhelming for the users. Ontologies or  LLMs could be used to further categorize the facets into logical groups, such as ``environment'' or ``personal attributes'', thereby keeping the interface cleaner. 

\paragraph{IR4: Enable fast and reliable assignment}
In our work, we used a very basic VLM for zero-shot assignment, trying to find a good trade-off between reliability and efficiency. Current VLMs are much more powerful when describing scene content~\cite{li_survey_2025}. However, they currently still require multiple seconds to describe a single image and are therefore not suitable for a real-time interactive EluDA system. In EluDA, this kind of in-depth analysis should, therefore, be restricted to a rather small set of potentially interesting images (based on a user model). Faster, approximate assignments should be performed at larger scales to improve data indexing. 

Furthermore, we used a pre-trained VLM without any fine-tuning. Few-shot learning approaches, e.g., leveraging adapters~\cite{gao_clip-adapter_2024}, could be a promising way to effectively fine-tune zero-shot models to more fine-grained problems. Few-shot learning requires careful interaction design for user intervention, balancing the required user effort (\textbf{IR2}) with reliability. % User intervention could be steered similarly to active learning approaches (see also \textbf{IR1}).

Even though we expect zero-shot and few-shot models to become more powerful and faster over time, we do not expect them to become fully reliable on a fine-grained level in a real-time setting in the near future. This calls for effective trust calibration methods~\cite{han_beyond_2020} and for adapting classic EDA approaches to account for underlying uncertainty. It is not straightforward to design an EDA system that promotes trust while not overwhelming the user with ``trust junk''~\cite{wall_trust_2024}. An EluDA system built on incomplete and potentially biased conceptualization, and on unreliable assignment of data items to concepts, needs to support fuzzy queries and filters, as well as uncertainty visualization techniques~\cite{kamal_recent_2021}. 

\section{Limitations}
 \label{sec:limitations}

Our study is formative and relies on low-fidelity interface prototypes. Our goal has been to provide participants with as much freedom as possible by mimicking a physical environment as if images had been printed. This leads to a couple of limitations: 

\begin{itemize}
    \item Due to the low-fidelity interface prototypes, conceptualization was not computationally \textbf{exploited to support EDA}. Opportunities, as discussed in Section ~\ref{sec:opportunities}, remain speculative and for future work. We expect that users would not perceive conceptualization as so laborious if they could see an immediate benefit for EDA. 
    \item The choice of the \textbf{conceptualization interface design} was guided by prior work, primarily in the field of visual analytics (see Sections~\ref{sec:knowledge-externalization} and \ref{sec:knowledge-exploitation}). Clearly, the choice of interface design can influence the users' pursued knowledge organization strategy (\textbf{RQ1}). Different conceptualization interface design variants are possible, and their impact on knowledge organization strategies should be investigated in the future. 
    \item To generate ground-truth assignments of externalized concepts to images via crowdsourcing, the study's \textbf{dataset size} was limited. This led to the clear limitation that users were, in fact, able to observe the entire dataset within a reasonable time frame. Some of our results, such as exhaustiveness and informativeness (\textbf{RQ3}), therefore cannot be generalized to EluDA settings where users cannot see the entire dataset or a representative subset thereof.
    \item Beyond data size, we restricted our study to a single, easily interpretable \textbf{data modality}: images. This choice allowed us to focus on evaluating the automatic assignments using a VLM. It also made data visualization straightforward. Other data modalities would necessitate different models and more sophisticated visualization techniques to provide users with an effective overview of the data.
    \item The non-expert \textbf{users} in our study were exclusively recruited from IT-related university programs, which introduces biases regarding their interaction strategies and their familiarity with externalization tools such as concept maps. Future case studies should therefore be conducted in domain expert settings.
    \item In our study, we focused on a general \textbf{task}. In practice, we are interested in expert analysts exploring domain-specific data, such as sensor data collected over a long time period, surveillance camera recordings, or collections of scientific documents. In the future, it will be necessary to investigate the extent to which the findings from our study apply to more complex and specialized EluDA scenarios.
    \item We used CLIP as a relatively basic \textbf{multimodal model} in our study, which is theoretically fast enough as a backbone for an interactive system but clearly does not achieve state-of-the-art performance in capturing visual concepts. The field is evolving rapidly, so more recent models may yield superior results without unacceptable latency in a real-time system.
\end{itemize}

\section{Conclusions}
\label{sec:conclusions}

EluDA adds a ``search for structure'' step to the exploratory analysis of large, unstructured data, where users conceptualize a structure and assign data items to it. In our formative study, conceptualizations are highly subjective and costly to construct. Users built faceted classifications bottom-up, aiming for exhaustive, data-grounded descriptions. 

To reduce the subjective burden of conceptualization, AI in EDA should maximize users’ exposure to diverse raw data rather than inferring or imposing concepts. 
This approach is supported by the fact that state-of-the-art multimodal models, which are efficient enough to be included in an interactive interface, are not reliable enough to autonomously assign data to highly specific concepts of interest to the user. 

From this, we derive key challenges and corresponding design requirements for future EluDA systems: Human-AI collaboration should help the user to 1) see as many diverse aspects of the data as possible, 2) describe the data exhaustively while respecting users' own perspectives and interests, 3) rapidly validate and iterate their conceptualization against the data, and 4) continuously query, filter, and visualize the data based on a potentially incomplete or biased conceptualization while being aware of uncertainties. 

\begin{acks}
This research was funded in whole or in part by the Austrian Science Fund (FWF) \href{https://doi.org/10.55776/P36453}{10.55776/P36453}. The authors acknowledge TU Wien Bibliothek for financial support through its Open Access Funding Programme.
\end{acks}

\bibliographystyle{ACM-Reference-Format}
\bibliography{bibliography}

@misc{ren_survey_2021,
    title = {A {Survey} of {Deep} {Active} {Learning}},
    urldate = {2023-08-03},
    publisher = {arXiv},
    author = {Ren, Pengzhen and Xiao, Yun and Chang, Xiaojun and Huang, Po-Yao and Li, Zhihui and Gupta, Brij B. and Chen, Xiaojiang and Wang, Xin},
    month = dec,
    year = {2021},
}

@article{hilasaca_grid-based_2024,
    title = {A {Grid}-{Based} {Method} for {Removing} {Overlaps} of {Dimensionality} {Reduction} {Scatterplot} {Layouts}},
    volume = {30},
    issn = {1941-0506},
    doi = {10.1109/TVCG.2023.3309941},
    number = {8},
    urldate = {2026-03-31},
    journal = {IEEE Transactions on Visualization and Computer Graphics},
    author = {Hilasaca, Gladys M. and Marcílio-Jr, Wilson E. and Eler, Danilo M. and Martins, Rafael M. and Paulovich, Fernando V.},
    month = aug,
    year = {2024},
    pages = {5733--5749},
}

@inproceedings{wall_warning_2017,
    title = {Warning, {Bias} {May} {Occur}: {A} {Proposed} {Approach} to {Detecting} {Cognitive} {Bias} in {Interactive} {Visual} {Analytics}},
    shorttitle = {Warning, {Bias} {May} {Occur}},
    doi = {10.1109/VAST.2017.8585669},
    publisher = {IEEE},
    address = {Phoenix, Arizona, USA},
    urldate = {2026-03-31},
    booktitle = {2017 {IEEE} {Conference} on {Visual} {Analytics} {Science} and {Technology} ({VAST})},
    author = {Wall, Emily and Blaha, Leslie M. and Franklin, Lyndsey and Endert, Alex},
    month = oct,
    year = {2017},
    pages = {104--115},
}

@article{sarvghad_visualizing_2017,
    title = {Visualizing {Dimension} {Coverage} to {Support} {Exploratory} {Analysis}},
    volume = {23},
    issn = {1941-0506},
    doi = {10.1109/TVCG.2016.2598466},
    number = {1},
    urldate = {2026-03-31},
    journal = {IEEE Transactions on Visualization and Computer Graphics},
    author = {Sarvghad, Ali and Tory, Melanie and Mahyar, Narges},
    month = jan,
    year = {2017},
    pages = {21--30},
}

@inproceedings{han_beyond_2020,
    title = {Beyond {Trust} {Building} — {Calibrating} {Trust} in {Visual} {Analytics}},
    doi = {10.1109/TREX51495.2020.00006},
    urldate = {2026-03-31},
    booktitle = {2020 {IEEE} {Workshop} on {TRust} and {EXpertise} in {Visual} {Analytics} ({TREX})},
    author = {Han, Wenkai and Schulz, Hans-Jörg},
    month = oct,
    year = {2020},
    publisher = {IEEE},
    address = {Salt Lake City, Utah, USA},
    pages = {9--15},
}

@inproceedings{wall_trust_2024,
    title = {Trust {Junk} and {Evil} {Knobs}: {Calibrating} {Trust} in {AI} {Visualization}},
    issn = {2165-8773},
    shorttitle = {Trust {Junk} and {Evil} {Knobs}},
    doi = {10.1109/PacificVis60374.2024.00012},
    urldate = {2026-03-31},
    booktitle = {2024 {IEEE} 17th {Pacific} {Visualization} {Conference} ({PacificVis})},
    author = {Wall, Emily and Matzen, Laura and El-Assady, Mennatallah and Masters, Peta and Hosseinpour, Helia and Endert, Alex and Borgo, Rita and Chau, Polo and Perer, Adam and Schupp, Harald and Strobelt, Hendrik and Padilla, Lace},
    month = apr,
    year = {2024},
    pages = {22--31},
    publisher = {IEEE},
    address = {Tokyo, Japan},
}

@article{kamal_recent_2021,
    title = {Recent advances and challenges in uncertainty visualization: a survey},
    volume = {24},
    issn = {1875-8975},
    shorttitle = {Recent advances and challenges in uncertainty visualization},
    doi = {10.1007/s12650-021-00755-1},
    language = {en},
    number = {5},
    urldate = {2026-03-31},
    journal = {Journal of Visualization},
    author = {Kamal, Aasim and Dhakal, Parashar and Javaid, Ahmad Y. and Devabhaktuni, Vijay K. and Kaur, Devinder and Zaientz, Jack and Marinier, Robert},
    month = oct,
    year = {2021},
    pages = {861--890},
}

@inproceedings{li_survey_2025,
    title = {A {Survey} of {State} of the {Art} {Large} {Vision} {Language} {Models}: {Alignment}, {Benchmark}, {Evaluations} and {Challenges}},
    issn = {2160-7516},
    shorttitle = {A {Survey} of {State} of the {Art} {Large} {Vision} {Language} {Models}},
    doi = {10.1109/CVPRW67362.2025.00147},
    urldate = {2026-04-07},
    booktitle = {2025 {IEEE}/{CVF} {Conference} on {Computer} {Vision} and {Pattern} {Recognition} {Workshops} ({CVPRW})},
    author = {Li, Zongxia and Wu, Xiyang and Du, Hongyang and Liu, Fuxiao and Nghiem, Huy and Shi, Guangyao},
    month = jun,
    year = {2025},
    publisher = {IEEE},
    address = {Nashville, Tennessee, USA},
    pages = {1578--1597},
}

@inproceedings{wenyin_mialbum_2000,
    address = {New York, NY, USA},
    series = {{MULTIMEDIA} '00},
    title = {{MiAlbum} - a system for home photo managemet using the semi-automatic image annotation approach},
    isbn = {978-1-58113-198-7},
    doi = {10.1145/354384.379011},
    urldate = {2026-04-07},
    booktitle = {Proceedings of the eighth {ACM} international conference on {Multimedia}},
    publisher = {Association for Computing Machinery},
    author = {Wenyin, Liu and Sun, Yanfeng and Zhang, Hongjiang},
    month = oct,
    year = {2000},
    pages = {479--480},
}

@book{hartwig_exploratory_1979,
    address = {Newbury Park, California},
    title = {Exploratory {Data} {Analysis}},
    isbn = {978-1-4129-8423-2},
    doi = {10.4135/9781412984232},
    language = {en},
    urldate = {2025-03-19},
    publisher = {SAGE Publications Inc.},
    author = {Hartwig, Frederick and Dearling, Brian E.},
    year = {1979},
}

@book{young_visual_2006,
    address = {Hoboken, NJ, USA},
    title = {Visual {Statistics}: {Seeing} {Data} with {Dynamic} {Interactive} {Graphics}},
    isbn = {978-1-118-16541-6},
    shorttitle = {Visual {Statistics}},
    language = {en},
    publisher = {John Wiley \& Sons},
    author = {Young, Forrest W. and Valero-Mora, Pedro M. and Friendly, Michael},
    month = jul,
    year = {2006},
}

@book{theus_interactive_2008,
    address = {1-4 Singer Street London},
    series = {Computer {Science} and {Data} {Analysi}},
    title = {Interactive {Graphics} for {Data} {Analysis}: {Principles} and {Examples}},
    isbn = {978-1-4200-1106-7},
    shorttitle = {Interactive {Graphics} for {Data} {Analysis}},
    language = {en},
    publisher = {CRC Press},
    author = {Theus, Martin and Urbanek, Simon},
    month = oct,
    year = {2008},
}

@article{blumberg_problem_2003,
    title = {The problem with unstructured data},
    volume = {13},
    number = {42-49},
    journal = {Dm Review},
    publisher = {Powell Publishing Inc},
    author = {Blumberg, Robert and Atre, Shaku},
    year = {2003},
    pages = {62},
}

@misc{wongsuphasawat_goals_2019,
    title = {Goals, {Process}, and {Challenges} of {Exploratory} {Data} {Analysis}: {An} {Interview} {Study}},
    shorttitle = {Goals, {Process}, and {Challenges} of {Exploratory} {Data} {Analysis}},
    doi = {10.48550/arXiv.1911.00568},
    language = {en},
    urldate = {2025-03-11},
    publisher = {arXiv},
    author = {Wongsuphasawat, Kanit and Liu, Yang and Heer, Jeffrey},
    month = nov,
    year = {2019},
}

@article{li_tetrahedral_2010,
    title = {A tetrahedral data model for unstructured data management},
    volume = {53},
    copyright = {http://www.springer.com/tdm},
    issn = {1674-733X, 1869-1919},
doi = {10.1007/s11432-010-4030-9},
    language = {en},
    number = {8},
    urldate = {2025-01-27},
    journal = {Science China Information Sciences},
    author = {Li, Wei and Lang, Bo},
    month = aug,
    year = {2010},
    pages = {1497--1510},
}

@article{buja_statistical_2009,
    title = {Statistical inference for exploratory data analysis and model diagnostics},
    volume = {367},
    issn = {1364-503X, 1471-2962},
    doi = {10.1098/rsta.2009.0120},
    language = {en},
    number = {1906},
    urldate = {2025-03-03},
    journal = {Philosophical Transactions of the Royal Society A: Mathematical, Physical and Engineering Sciences},
    author = {Buja, Andreas and Cook, Dianne and Hofmann, Heike and Lawrence, Michael and Lee, Eun-Kyung and Swayne, Deborah F. and Wickham, Hadley},
    month = nov,
    year = {2009},
    pages = {4361--4383},
}

@article{stolte_polaris_2002,
    title = {Polaris: a system for query, analysis, and visualization of multidimensional relational databases},
    volume = {8},
    doi = {10.1109/2945.981851},
    number = {1},
    journal = {IEEE Transactions on Visualization and Computer Graphics},
    author = {Stolte, C. and Tang, D. and Hanrahan, P.},
    year = {2002},
    pages = {52--65},
}

@article{hai_data_2023,
    title = {Data {Lakes}: {A} {Survey} of {Functions} and {Systems}},
    volume = {35},
    issn = {1558-2191},
    shorttitle = {Data {Lakes}},
    doi = {10.1109/TKDE.2023.3270101},
    number = {12},
    urldate = {2026-01-26},
    journal = {IEEE Transactions on Knowledge and Data Engineering},
    author = {Hai, Rihan and Koutras, Christos and Quix, Christoph and Jarke, Matthias},
    month = dec,
    year = {2023},
    pages = {12571--12590},
}

@inproceedings{nargesian_organizing_2020,
    address = {New York, NY, USA},
    series = {{SIGMOD} '20},
    title = {Organizing {Data} {Lakes} for {Navigation}},
    isbn = {978-1-4503-6735-6},
    doi = {10.1145/3318464.3380605},
    urldate = {2026-01-26},
    booktitle = {Proceedings of the 2020 {ACM} {SIGMOD} {International} {Conference} on {Management} of {Data}},
    publisher = {Association for Computing Machinery},
    author = {Nargesian, Fatemeh and Pu, Ken Q. and Zhu, Erkang and Ghadiri Bashardoost, Bahar and Miller, Renée J.},
    month = may,
    year = {2020},
    pages = {1939--1950},
}

@article{fisher_use_1936,
    title = {The {Use} of {Multiple} {Measurements} in {Taxonomic} {Problems}},
    volume = {7},
    issn = {2050-1439},
    doi = {10.1111/j.1469-1809.1936.tb02137.x},
    language = {en},
    number = {2},
    urldate = {2025-08-22},
    journal = {Annals of Eugenics},
    author = {Fisher, R. A.},
    year = {1936},

    pages = {179--188},
}

@inproceedings{krizhevsky_imagenet_2012,
    address = {Lake Tahoe, Nevada},
    title = {{ImageNet} {Classification} with {Deep} {Convolutional} {Neural} {Networks}},
    volume = {25},
    urldate = {2025-08-22},
    booktitle = {Advances in {Neural} {Information} {Processing} {Systems}},
    publisher = {Curran Associates, Inc.},
    author = {Krizhevsky, Alex and Sutskever, Ilya and Hinton, Geoffrey E},
    year = {2012},
    pages = {1097--1105},
}

@inproceedings{he_deep_2016,
    address = {Las Vegas, NV, USA},
    title = {Deep {Residual} {Learning} for {Image} {Recognition}},
    isbn = {978-1-4673-8851-1},
    doi = {10.1109/CVPR.2016.90},
    language = {en},
    urldate = {2025-08-22},
    booktitle = {2016 {IEEE} {Conference} on {Computer} {Vision} and {Pattern} {Recognition} ({CVPR})},
    publisher = {IEEE},
    author = {He, Kaiming and Zhang, Xiangyu and Ren, Shaoqing and Sun, Jian},
    month = jun,
    year = {2016},
    pages = {770--778},
}

@article{kowsari_text_2019,
    title = {Text {Classification} {Algorithms}: {A} {Survey}},
    volume = {10},
    copyright = {http://creativecommons.org/licenses/by/3.0/},
    issn = {2078-2489},
    shorttitle = {Text {Classification} {Algorithms}},
    doi = {10.3390/info10040150},
    language = {en},
    number = {4},
    urldate = {2025-09-03},
    journal = {Information},
    publisher = {Multidisciplinary Digital Publishing Institute},
    author = {Kowsari, Kamran and Jafari Meimandi, Kiana and Heidarysafa, Mojtaba and Mendu, Sanjana and Barnes, Laura and Brown, Donald},
    month = apr,
    year = {2019},
    pages = {150},
}

@article{jia_towards_2022,
    title = {Towards {Visual} {Explainable} {Active} {Learning} for {Zero}-{Shot} {Classification}},
    volume = {28},
    issn = {1941-0506},
    doi = {10.1109/TVCG.2021.3114793},
    number = {1},
    journal = {IEEE Transactions on Visualization and Computer Graphics},
    author = {Jia, Shichao and Li, Zeyu and Chen, Nuo and Zhang, Jiawan},
    month = jan,
    year = {2022},
    pages = {791--801},
}

@article{bernard_vial_2018,
    title = {{VIAL}: a unified process for visual interactive labeling},
    volume = {34},
    issn = {1432-2315},
    shorttitle = {{VIAL}},
    doi = {10.1007/s00371-018-1500-3},
    language = {en},
    number = {9},
    urldate = {2025-03-25},
    journal = {The Visual Computer},
    author = {Bernard, Jürgen and Zeppelzauer, Matthias and Sedlmair, Michael and Aigner, Wolfgang},
    month = sep,
    year = {2018},
    pages = {1189--1207},
}

@misc{radford_learning_2021,
    title = {Learning {Transferable} {Visual} {Models} {From} {Natural} {Language} {Supervision}},
    language = {en},
    urldate = {2024-05-13},
    publisher = {arXiv},
    author = {Radford, Alec and Kim, Jong Wook and Hallacy, Chris and Ramesh, Aditya and Goh, Gabriel and Agarwal, Sandhini and Sastry, Girish and Askell, Amanda and Mishkin, Pamela and Clark, Jack and Krueger, Gretchen and Sutskever, Ilya},
    month = feb,
    year = {2021},
}

@article{cooper_rethinking_2025,
    title = {Rethinking {VLMs} and {LLMs} for image classification},
    volume = {15},
    copyright = {2025 The Author(s)},
    issn = {2045-2322},
doi = {10.1038/s41598-025-04384-8},
    language = {en},
    number = {1},
    urldate = {2026-01-22},
    journal = {Scientific Reports},
    publisher = {Nature Publishing Group},
    author = {Cooper, Avi and Kato, Keizo and Shih, Chia-Hsien and Yamane, Hiroaki and Vinken, Kasper and Takemoto, Kentaro and Sunagawa, Taro and Yeh, Hao-Wei and Yamanaka, Jin and Mason, Ian and Boix, Xavier},
    month = jun,
    year = {2025},
    pages = {19692},
}

@article{eschner_interactive_2025,
    title = {Interactive {Discovery} and {Exploration} of {Visual} {Bias} in {Generative} {Text}-to-{Image} {Models}},
    volume = {44},
    copyright = {© 2025 The Author(s). Computer Graphics Forum published by Eurographics - The European Association for Computer Graphics and John Wiley \& Sons Ltd.},
    issn = {1467-8659},
doi = {10.1111/cgf.70135},
    language = {en},
    number = {3},
    urldate = {2025-09-01},
    journal = {Computer Graphics Forum},
    author = {Eschner, Johannes and Labadie-Tamayo, Roberto and Zeppelzauer, Matthias and Waldner, Manuela},
    year = {2025},
    pages = {e70135},
}

@article{wagner_kavagait_2019,
    title = {{KAVAGait}: {Knowledge}-{Assisted} {Visual} {Analytics} for {Clinical} {Gait} {Analysis}},
    volume = {25},
    copyright = {https://creativecommons.org/licenses/by/3.0/legalcode},
    issn = {1077-2626, 1941-0506, 2160-9306},
    shorttitle = {{KAVAGait}},
doi = {10.1109/TVCG.2017.2785271},
    language = {en},
    number = {3},
    urldate = {2025-01-28},
    journal = {IEEE Transactions on Visualization and Computer Graphics},
    author = {Wagner, Markus and Slijepcevic, Djordje and Horsak, Brian and Rind, Alexander and Zeppelzauer, Matthias and Aigner, Wolfgang},
    month = mar,
    year = {2019},
    pages = {1528--1542},
}

@article{li_incorporation_2022,
    title = {Incorporation of {Human} {Knowledge} {Into} {Data} {Embeddings} to {Improve} {Pattern} {Significance} and {Interpretability}},
    volume = {29},
    copyright = {https://ieeexplore.ieee.org/Xplorehelp/downloads/license-information/IEEE.html},
    issn = {1077-2626, 1941-0506, 2160-9306},
doi = {10.1109/TVCG.2022.3209382},
    language = {en},
    number = {1},
    urldate = {2025-01-28},
    journal = {IEEE Transactions on Visualization and Computer Graphics},
    author = {Li, Jie and Zhou, Chun-qi},
    year = {2022},
    pages = {1--11},
}

@article{wang_kmtlabeler_2024,
    title = {{KMTLabeler}: {An} {Interactive} {Knowledge}-{Assisted} {Labeling} {Tool} for {Medical} {Text} {Classification}},
    issn = {1941-0506},
    shorttitle = {{KMTLabeler}},
doi = {10.1109/TVCG.2024.3406387},
    urldate = {2024-10-15},
    journal = {IEEE Transactions on Visualization and Computer Graphics},
    author = {Wang, He and Ouyang, Yang and Wu, Yuchen and Jiang, Chang and Jin, Lixia and Cao, Yuanwu and Li, Quan},
    year = {2024},
    volume = {39},
    pages = {1--18},
}

@incollection{nonaka_knowledge-creating_1998,
    address = {Boston},
    title = {The {Knowledge}-{Creating} {Company}},
    isbn = {978-0-7506-7009-8},
    booktitle = {The {Economic} {Impact} of {Knowledge}},
    publisher = {Butterworth-Heinemann},
    author = {Nonaka, Ikujiro},
    year = {1998},

    pages = {175--187},
}

@inproceedings{federico_role_2017,
    address = {Phoenix, AZ},
    title = {The {Role} of {Explicit} {Knowledge}: {A} {Conceptual} {Model} of {Knowledge}-{Assisted} {Visual} {Analytics}},
    isbn = {978-1-5386-3163-8},
    shorttitle = {The {Role} of {Explicit} {Knowledge}},
doi = {10.1109/VAST.2017.8585498},
    language = {en},
    urldate = {2023-08-04},
    booktitle = {2017 {IEEE} {Conference} on {Visual} {Analytics} {Science} and {Technology} ({VAST})},
    publisher = {IEEE},
    author = {Federico, Paolo and Wagner, Markus and Rind, Alexander and Amor-Amoros, Albert and Miksch, Silvia and Aigner, Wolfgang},
    month = oct,
    year = {2017},
    pages = {92--103},
}

@article{hogan_knowledge_2022,
    title = {Knowledge {Graphs}},
    volume = {54},
    issn = {0360-0300, 1557-7341},
doi = {10.1145/3447772},
    language = {en},
    number = {4},
    urldate = {2025-03-11},
    journal = {ACM Computing Surveys},
    author = {Hogan, Aidan and Blomqvist, Eva and Cochez, Michael and D’amato, Claudia and Melo, Gerard De and Gutierrez, Claudio and Kirrane, Sabrina and Gayo, José Emilio Labra and Navigli, Roberto and Neumaier, Sebastian and Ngomo, Axel-Cyrille Ngonga and Polleres, Axel and Rashid, Sabbir M. and Rula, Anisa and Schmelzeisen, Lukas and Sequeda, Juan and Staab, Steffen and Zimmermann, Antoine},
    month = may,
    year = {2022},
    pages = {1--37},
}

@article{nonaka_dynamic_1994,
    title = {A {Dynamic} {Theory} of {Organizational} {Knowledge} {Creation}},
    volume = {5},
    issn = {1047-7039},
    number = {1},
    urldate = {2025-08-26},
    journal = {Organization Science},
    publisher = {INFORMS},
    author = {Nonaka, Ikujiro},
    year = {1994},
    pages = {14--37},
}

@book{broughton_essential_2015,
    address = {London, United Kingdom},
    title = {Essential {Classification}},
    isbn = {978-1-78330-031-0},
publisher = {Facet Publishing},
    author = {Broughton, V.},
    year = {2015},
}

@article{dawson_need_2006,
    title = {The need for a faceted classification as the basis of all methods of information retrieval},
    volume = {58},
    copyright = {http://www.emeraldinsight.com/page/tdm},
    issn = {0001-253X},
doi = {10.1108/00012530610648671},
    language = {en},
    number = {1/2},
    urldate = {2025-03-13},
    journal = {Aslib Proceedings},
    author = {Broughton, Vanda},
    editor = {Dawson, Andy},
    month = jan,
    year = {2006},
    pages = {49--72},
}

@article{krotzsch_editorial_2016,
    title = {Editorial},
    volume = {37-38},
    issn = {15708268},
doi = {10.1016/j.websem.2016.04.002},
    language = {en},
    urldate = {2025-03-20},
    journal = {Journal of Web Semantics},
    author = {Krötzsch, Markus and Weikum, Gerhard},
    month = mar,
    year = {2016},
    pages = {53--54},
}

@incollection{fellbaum_wordnet_2010,
    address = {Dordrecht},
    title = {{WordNet}},
    isbn = {978-90-481-8847-5},
doi = {10.1007/978-90-481-8847-5_10},
    language = {en},
    urldate = {2025-03-20},
    booktitle = {Theory and {Applications} of {Ontology}: {Computer} {Applications}},
    publisher = {Springer Netherlands},
    author = {Fellbaum, Christiane},
    editor = {Poli, Roberto and Healy, Michael and Kameas, Achilles},
    year = {2010},
    pages = {231--243},
}

@article{vrandecic_wikidata_2014,
    title = {Wikidata: a free collaborative knowledgebase},
    volume = {57},
    issn = {0001-0782, 1557-7317},
    shorttitle = {Wikidata},
doi = {10.1145/2629489},
    language = {en},
    number = {10},
    urldate = {2025-03-20},
    journal = {Communications of the ACM},
    author = {Vrandečić, Denny and Krötzsch, Markus},
    month = sep,
    year = {2014},
    pages = {78--85},
}

@article{jacob_classification_2004,
    title = {Classification and {Categorization}: {A} {Difference} that {Makes} a {Difference}},
    volume = {52},
    copyright = {Copyright owned by Board of Trustees of the University of Illinois. 2004.},
    issn = {0024-2594},
    shorttitle = {Classification and {Categorization}},
    language = {en},
    number = {3},
    urldate = {2025-03-19},
    journal = {Library Trends},
    publisher = {Graduate School of Library and Information Science. University of Illinois at Urbana-Champaign.},
    author = {Jacob, Elin K.},
    year = {2004},
    pages = {515--540},
}

@article{mills_faceted_2004,
    title = {Faceted {Classification} and {Logical} {Division} in {Information} {Retrieval}},
    volume = {52},
journal = {Libr. Trends},
    author = {Mills, Jack},
    year = {2004},
    pages = {541--570},
}

@inproceedings{prieto-diaz_faceted_2003,
    address = {Las Vegas, NV, USA},
    title = {A faceted approach to building ontologies},
doi = {10.1109/IRI.2003.1251451},
    urldate = {2025-09-08},
    booktitle = {Proceedings {Fifth} {IEEE} {Workshop} on {Mobile} {Computing} {Systems} and {Applications}},
    publisher = {IEEE},
    author = {Prieto-Diaz, R.},
    month = oct,
    year = {2003},
    pages = {458--465},
}

@book{ranganathan_colon_1933,
    address = {Madras, London},
    series = {Publication series ({Madras} {Library} {Association})},
    title = {Colon classification},
    language = {eng},
    publisher = {The Madras Library Association ; Goldston},
    author = {Ranganathan, S. R.},
    year = {1933},
}

@article{hjorland_facet_2013,
    title = {Facet analysis: {The} logical approach to knowledge organization},
    volume = {49},
    copyright = {https://www.elsevier.com/tdm/userlicense/1.0/},
    issn = {03064573},
    shorttitle = {Facet analysis},
doi = {10.1016/j.ipm.2012.10.001},
    language = {en},
    number = {2},
    urldate = {2025-03-13},
    journal = {Information Processing \& Management},
    author = {Hjørland, Birger},
    month = mar,
    year = {2013},
    pages = {545--557},
}

@article{gupta_survey_2010,
    title = {Survey on social tagging techniques},
    volume = {12},
    number = {1},
    journal = {ACM SIGKDD Explorations Newsletter},
    publisher = {ACM New York, NY, USA},
    author = {Gupta, Manish and Li, Rui and Yin, Zhijun and Han, Jiawei},
    year = {2010},
    pages = {58--72},
}

@book{peters_folksonomies_2009,
    address = {USA},
    edition = {1st},
    title = {Folksonomies. {Indexing} and {Retrieval} in {Web} 2.0},
    isbn = {978-3-598-25179-5},
    publisher = {Walter de Gruyter \& Co.},
    author = {Peters, Isabella},
    month = sep,
    year = {2009},
}

@inproceedings{halpin_complex_2007,
    address = {New York, NY, USA},
    series = {{WWW} '07},
    title = {The complex dynamics of collaborative tagging},
    isbn = {978-1-59593-654-7},
doi = {10.1145/1242572.1242602},
    urldate = {2025-09-08},
    booktitle = {Proceedings of the 16th international conference on {World} {Wide} {Web}},
    publisher = {Association for Computing Machinery},
    author = {Halpin, Harry and Robu, Valentin and Shepherd, Hana},
    month = may,
    year = {2007},
    pages = {211--220},
}

@article{chen_review_2020,
    title = {A review: {Knowledge} reasoning over knowledge graph},
    volume = {141},
    issn = {09574174},
    shorttitle = {A review},
doi = {10.1016/j.eswa.2019.112948},
    language = {en},
    urldate = {2025-03-11},
    journal = {Expert Systems with Applications},
    author = {Chen, Xiaojun and Jia, Shengbin and Xiang, Yang},
    month = mar,
    year = {2020},
    pages = {112948},
}

@article{rosch_basic_1976,
    title = {Basic objects in natural categories},
    volume = {8},
    copyright = {https://www.elsevier.com/tdm/userlicense/1.0/},
    issn = {00100285},
doi = {10.1016/0010-0285(76)90013-X},
    language = {en},
    number = {3},
    urldate = {2025-01-29},
    journal = {Cognitive Psychology},
    author = {Rosch, Eleanor and Mervis, Carolyn B and Gray, Wayne D and Johnson, David M and Boyes-Braem, Penny},
    month = jul,
    year = {1976},
    pages = {382--439},
}

@article{lohfink_knowledge_2022,
    title = {Knowledge {Rocks}: {Adding} {Knowledge} {Assistance} to {Visualization} {Systems}},
    volume = {28},
    issn = {1941-0506},
    shorttitle = {Knowledge {Rocks}},
doi = {10.1109/TVCG.2021.3114687},
    number = {1},
    urldate = {2024-06-06},
    journal = {IEEE Transactions on Visualization and Computer Graphics},
    author = {Lohfink, Anna-Pia and Duque Anton, Simon D. and Leitte, Heike and Garth, Christoph},
    month = jan,
    year = {2022},
    pages = {1117--1127},
}

@article{novak_theory_2006,
    title = {The {Theory} {Underlying} {Concept} {Maps} and {How} to {Construct} and {Use} {Them}},
    volume = {2006},
    language = {en},
    number = {01},
    journal = {Florida Institute for Human and Machine Cognition},
    author = {Novak, Joseph D and Cañas, Alberto J},
    year = {2006},
    pages = {1--36},
}

@article{watson_assessing_2016,
    title = {Assessing {Conceptual} {Knowledge} {Using} {Three} {Concept} {Map} {Scoring} {Methods}},
    volume = {105},
    copyright = {http://onlinelibrary.wiley.com/termsAndConditions\#vor},
    issn = {1069-4730, 2168-9830},
doi = {10.1002/jee.20111},
    language = {en},
    number = {1},
    urldate = {2025-01-29},
    journal = {Journal of Engineering Education},
    author = {Watson, Mary Katherine and Pelkey, Joshua and Noyes, Caroline R. and Rodgers, Michael O.},
    month = jan,
    year = {2016},
    pages = {118--146},
}

@article{starr_concept_2013,
    title = {Concept maps as the first step in an ontology construction method},
    volume = {38},
    issn = {03064379},
doi = {10.1016/j.is.2012.05.010},
    language = {en},
    number = {5},
    urldate = {2025-01-29},
    journal = {Information Systems},
    author = {Starr, Rodrigo Rizzi and Parente De Oliveira, José Maria},
    month = jul,
    year = {2013},
    pages = {771--783},
}

@article{zhao_supporting_2018,
    title = {Supporting {Handoff} in {Asynchronous} {Collaborative} {Sensemaking} {Using} {Knowledge}-{Transfer} {Graphs}},
    volume = {24},
    issn = {1941-0506},
doi = {10.1109/TVCG.2017.2745279},
    number = {1},
    urldate = {2025-03-19},
    journal = {IEEE Transactions on Visualization and Computer Graphics},
    author = {Zhao, Jian and Glueck, Michael and Isenberg, Petra and Chevalier, Fanny and Khan, Azam},
    month = jan,
    year = {2018},
    pages = {340--350},
}

@inproceedings{mahyar_closer_2010,
    title = {A closer look at note taking in the co-located collaborative visual analytics process},
doi = {10.1109/VAST.2010.5652879},
    urldate = {2024-10-01},
    booktitle = {2010 {IEEE} {Symposium} on {Visual} {Analytics} {Science} and {Technology}},
    author = {Mahyar, Narges and Sarvghad, Ali and Tory, Melanie},
    month = oct,
    year = {2010},
    publisher = {IEEE},
    address = {Salt Lake City, Utah, USA},
    pages = {171--178},
}

@article{waldner_linking_2021,
    title = {Linking unstructured evidence to structured observations},
    volume = {20},
    issn = {1473-8716},
doi = {10.1177/1473871620986249},
    language = {en},
    number = {1},
    urldate = {2023-08-29},
    journal = {Information Visualization},
    publisher = {SAGE Publications},
    author = {Waldner, Manuela and Geymayer, Thomas and Schmalstieg, Dieter and Sedlmair, Michael},
    month = jan,
    year = {2021},
    pages = {47--65},
}

@article{mahyar_supporting_2014,
    title = {Supporting communication and coordination in collaborative sensemaking},
    volume = {20},
    number = {12},
    journal = {IEEE transactions on visualization and computer graphics},
    publisher = {IEEE},
    author = {Mahyar, Narges and Tory, Melanie},
    year = {2014},
    pages = {1633--1642},
}

@article{kirsh_intelligent_1995,
    series = {Computational {Research} on {Interaction} and {Agency}, {Part} 2},
    title = {The intelligent use of space},
    volume = {73},
    issn = {0004-3702},
doi = {10.1016/0004-3702(94)00017-U},
    number = {1},
    urldate = {2025-03-19},
    journal = {Artificial Intelligence},
    author = {Kirsh, David},
    month = feb,
    year = {1995},
    pages = {31--68},
}

@article{malone_how_1983,
    title = {How do people organize their desks?: {Implications} for the design of office information systems},
    volume = {1},
    issn = {1046-8188, 1558-2868},
    shorttitle = {How do people organize their desks?},
doi = {10.1145/357423.357430},
    language = {en},
    number = {1},
    urldate = {2025-03-19},
    journal = {ACM Transactions on Information Systems},
    author = {Malone, Thomas W.},
    month = jan,
    year = {1983},
    pages = {99--112},
}

@inproceedings{mander_pile_1992,
    address = {Monterey, California, United States},
    title = {A “pile” metaphor for supporting casual organization of information},
    isbn = {978-0-89791-513-7},
doi = {10.1145/142750.143055},
    language = {en},
    urldate = {2025-03-19},
    booktitle = {Proceedings of the {SIGCHI} conference on {Human} factors in computing systems  - {CHI} '92},
    publisher = {ACM Press},
    author = {Mander, Richard and Salomon, Gitta and Wong, Yin Yin},
    year = {1992},
    pages = {627--634},
}

@inproceedings{andrews_space_2010,
    address = {Atlanta Georgia USA},
    title = {Space to think: large high-resolution displays for sensemaking},
    isbn = {978-1-60558-929-9},
    shorttitle = {Space to think},
doi = {10.1145/1753326.1753336},
    language = {en},
    urldate = {2025-01-28},
    booktitle = {Proceedings of the {SIGCHI} {Conference} on {Human} {Factors} in {Computing} {Systems}},
    publisher = {ACM},
    author = {Andrews, Christopher and Endert, Alex and North, Chris},
    month = apr,
    year = {2010},
    pages = {55--64},
}

@inproceedings{cetin_visual_2018,
    address = {Konstanz},
    title = {Visual {Analytics} on {Large} {Displays}: {Exploring} {User} {Spatialization} and {How} {Size} and {Resolution} {Affect} {Task} {Performance}},
    isbn = {978-1-5386-9194-6},
    shorttitle = {Visual {Analytics} on {Large} {Displays}},
doi = {10.1109/BDVA.2018.8534027},
    language = {en},
    urldate = {2025-03-19},
    booktitle = {2018 {International} {Symposium} on {Big} {Data} {Visual} and {Immersive} {Analytics} ({BDVA})},
    publisher = {IEEE},
    author = {Cetin, Gokhan and Stuerzlinger, Wolfgang and Dill, John},
    month = oct,
    year = {2018},
    pages = {1--10},
}

@inproceedings{endert_semantic_2012,
    address = {New York, NY, USA},
    series = {{CHI} '12},
    title = {Semantic interaction for visual text analytics},
    isbn = {978-1-4503-1015-4},
doi = {10.1145/2207676.2207741},
    urldate = {2023-09-22},
    booktitle = {Proceedings of the {SIGCHI} {Conference} on {Human} {Factors} in {Computing} {Systems}},
    publisher = {Association for Computing Machinery},
    author = {Endert, Alex and Fiaux, Patrick and North, Chris},
    year = {2012},
    pages = {473--482},
}

@incollection{buchanan_dendral_1981,
    address = {Stanford, CA, United States},
    title = {Dendral and {Meta}-{Dendral}},
    copyright = {https://www.elsevier.com/tdm/userlicense/1.0/},
    isbn = {978-0-934613-03-3},
doi = {10.1016/B978-0-934613-03-3.50026-X},
    language = {en},
    urldate = {2025-08-26},
    booktitle = {Readings in {Artificial} {Intelligence}},
    publisher = {Elsevier},
    author = {Buchanan, Bruce G. and Feigenbaum, Edward A.},
    year = {1981},
    pages = {313--322},
}

@article{peng_knowledge_2023,
    title = {Knowledge {Graphs}: {Opportunities} and {Challenges}},
    volume = {56},
    issn = {0269-2821, 1573-7462},
    shorttitle = {Knowledge {Graphs}},
doi = {10.1007/s10462-023-10465-9},
    language = {en},
    number = {11},
    urldate = {2025-03-19},
    journal = {Artificial Intelligence Review},
    author = {Peng, Ciyuan and Xia, Feng and Naseriparsa, Mehdi and Osborne, Francesco},
    month = nov,
    year = {2023},
    pages = {13071--13102},
}

@inproceedings{ge_exnav_2020,
    address = {Atlanta, GA, USA},
    title = {{ExNav}: {An} {Interactive} {Big} {Data} {Exploration} {Framework} for {Big} {Unstructured} {Data}},
    copyright = {https://ieeexplore.ieee.org/Xplorehelp/downloads/license-information/IEEE.html},
    isbn = {978-1-72816-251-5},
    shorttitle = {{ExNav}},
doi = {10.1109/BigData50022.2020.9377741},
    language = {en},
    urldate = {2025-01-28},
    booktitle = {2020 {IEEE} {International} {Conference} on {Big} {Data} ({Big} {Data})},
    publisher = {IEEE},
    author = {Ge, Xiaoyu and Zhang, Xiaozhong and Chrysanthis, Panos K.},
    month = dec,
    year = {2020},
    pages = {503--512},
}

@inproceedings{guo_expert---loop_2016,
    address = {Cham},
    title = {An {Expert}-in-the-loop {Paradigm} for {Learning} {Medical} {Image} {Grouping}},
    isbn = {978-3-319-31753-3},
    doi = {10.1007/978-3-319-31753-3_38},
    language = {en},
    booktitle = {Advances in {Knowledge} {Discovery} and {Data} {Mining}},
    publisher = {Springer International Publishing},
    author = {Guo, Xuan and Yu, Qi and Li, Rui and Alm, Cecilia Ovesdotter and Calvelli, Cara and Shi, Pengcheng and Haake, Anne},
    editor = {Bailey, James and Khan, Latifur and Washio, Takashi and Dobbie, Gill and Huang, Joshua Zhexue and Wang, Ruili},
    year = {2016},
    pages = {477--488},
}

@inproceedings{setlur_supporting_2025,
    address = {Berlin Germany},
    title = {Supporting {Human}-{Centric} {Data} {Exploration} {Through} {Semantics} and {Natural} {Language} {Interaction}},
    isbn = {979-8-4007-1564-8},
doi = {10.1145/3722212.3725628},
    language = {en},
    urldate = {2026-01-20},
    booktitle = {Companion of the 2025 {International} {Conference} on {Management} of {Data}},
    publisher = {ACM},
    author = {Setlur, Vidya},
    month = jun,
    year = {2025},
    pages = {851--854},
}

@inproceedings{vasu_mobileclip_2024,
    address = {Seattle, WA, USA},
    title = {{MobileCLIP}: {Fast} {Image}-{Text} {Models} through {Multi}-{Modal} {Reinforced} {Training}},
    copyright = {https://doi.org/10.15223/policy-029},
    isbn = {979-8-3503-5300-6},
    shorttitle = {{MobileCLIP}},
doi = {10.1109/CVPR52733.2024.01511},
    language = {en},
    urldate = {2026-03-13},
    booktitle = {2024 {IEEE}/{CVF} {Conference} on {Computer} {Vision} and {Pattern} {Recognition} ({CVPR})},
    publisher = {IEEE},
    author = {Vasu, Pavan Kumar Anasosalu and Pouransari, Hadi and Faghri, Fartash and Vemulapalli, Raviteja and Tuzel, Oncel},
    month = jun,
    year = {2024},
    pages = {15963--15974},
}

@inproceedings{zhong_clip4retrofit_2025,
    title = {{Clip4Retrofit}: {Enabling} {Real}-{Time} {Image} {Labeling} on {Edge} {Devices} via {Cross}-{Architecture} {CLIP} {Distillation}},
    shorttitle = {{Clip4Retrofit}},
language = {en},
    urldate = {2026-03-13},
    author = {Zhong, Li and Ghazal, Ahmed and Wan, Jun-Jun and Zilly, Frederik and Mackens, Patrick and Vollrath, Joachim and Coseriu, Bogdan},
    year = {2025},
    pages = {3868--3876},
    booktitle = {Proceedings of the IEEE/CVF Conference on Computer Vision and Pattern Recognition},
    address = {Nashville, Tennessee, USA},
    publisher = {IEEE},
}

@inproceedings{yang_mobileviclip_2025,
    title = {{MobileViCLIP}: {An} {Efficient} {Video}-{Text} {Model} for {Mobile} {Devices}},
    shorttitle = {{MobileViCLIP}},
language = {en},
    urldate = {2026-03-16},
    author = {Yang, Min and Jia, Zihan and Dai, Zhilin and Guo, Sheng and Wang, Limin},
    year = {2025},
    pages = {20824--20835},
    booktitle = {Proceedings of the IEEE/CVF Conference on Computer Vision and Pattern Recognition},
    address = {Nashville, Tennessee, USA},
    publisher = {IEEE},
}

@misc{podell_sdxl_2023,
    title = {{SDXL}: {Improving} {Latent} {Diffusion} {Models} for {High}-{Resolution} {Image} {Synthesis}},
    shorttitle = {{SDXL}},
doi = {10.48550/arXiv.2307.01952},
    language = {en},
    urldate = {2025-01-29},
    publisher = {arXiv},
    author = {Podell, Dustin and English, Zion and Lacey, Kyle and Blattmann, Andreas and Dockhorn, Tim and Müller, Jonas and Penna, Joe and Rombach, Robin},
    month = jul,
    year = {2023},
}

@misc{bureau_of_labor_statistics_labor_2023,
    title = {Labor {Force} {Statistics} from the {Current} {Population} {Survey}},
    url = {https://www.bls.gov/cps/cpsaat11.htm},
    language = {en},
    urldate = {2025-01-29},
    journal = {Bureau of Labor Statistics},
    author = {Bureau of Labor Statistics},
    year = {2023},
}

@incollection{vedaldi_simulating_2020,
    address = {Cham},
    title = {Simulating {Content} {Consistent} {Vehicle} {Datasets} with {Attribute} {Descent}},
    volume = {12351},
    isbn = {978-3-030-58538-9 978-3-030-58539-6},
doi = {10.1007/978-3-030-58539-6_46},
    language = {en},
    urldate = {2025-01-29},
    booktitle = {Computer {Vision} – {ECCV} 2020},
    publisher = {Springer International Publishing},
    author = {Yao, Yue and Zheng, Liang and Yang, Xiaodong and Naphade, Milind and Gedeon, Tom},
    editor = {Vedaldi, Andrea and Bischof, Horst and Brox, Thomas and Frahm, Jan-Michael},
    year = {2020},

    pages = {775--791},
}

@article{al-diban_comparison_2011,
    chapter = {Full Length Articles},
    title = {Comparison of {Two} {Analysis} {Approaches} for {Measuring} {Externalized} {Mental} {Models}},
    volume = {14},
    copyright = {© 2011. Notwithstanding the ProQuest Terms and Conditions, you may use this content in accordance with the associated terms available at https://www.j-ets.net/ETS/guide.html},
    issn = {11763647},
    language = {English},
    number = {2},
    urldate = {2025-03-19},
    journal = {Journal of Educational Technology \& Society},
    publisher = {International Forum of Educational Technology \& Society},
    author = {Al-Diban, Sabine and Ifenthaler, Dirk},
    year = {2011},
    pages = {16--30},
}

@inproceedings{tatu_combining_2009,
    address = {Atlantic City, NJ, USA},
    title = {Combining automated analysis and visualization techniques for effective exploration of high-dimensional data},
    isbn = {978-1-4244-5283-5},
doi = {10.1109/VAST.2009.5332628},
    language = {en},
    urldate = {2025-03-19},
    booktitle = {2009 {IEEE} {Symposium} on {Visual} {Analytics} {Science} and {Technology}},
    publisher = {IEEE},
    author = {Tatu, Andrada and Albuquerque, Georgia and Eisemann, Martin and Schneidewind, Jorn and Theisel, Holger and Magnork, Marcus and Keim, Daniel},
    year = {2009},
    pages = {59--66},
}

@article{sips_selecting_2009,
    title = {Selecting good views of high‐dimensional data using class consistency},
    volume = {28},
    issn = {0167-7055, 1467-8659},
doi = {10.1111/j.1467-8659.2009.01467.x},
    language = {en},
    number = {3},
    urldate = {2025-03-19},
    journal = {Computer Graphics Forum},
    author = {Sips, Mike and Neubert, Boris and Lewis, John P. and Hanrahan, Pat},
    month = jun,
    year = {2009},
    pages = {831--838},
}

@inproceedings{aupetit_sepme_2016,
    address = {Taipei, Taiwan},
    title = {{SepMe}: 2002 {New} visual separation measures},
    issn = {2165-8773},
    shorttitle = {{SepMe}},
    doi = {10.1109/PACIFICVIS.2016.7465244},
    urldate = {2024-01-18},
    booktitle = {2016 {IEEE} {Pacific} {Visualization} {Symposium} ({PacificVis})},
    author = {Aupetit, Michael and Sedlmair, Michael},
    publisher = {IEEE},
    month = apr,
    year = {2016},
    pages = {1--8},
}

@article{richmond_set_2014,
    title = {A set of guidelines for the consistent assessment of concept maps},
    volume = {30},
    issn = {0949-149X},
    language = {English (US)},
    number = {5},
    journal = {International Journal of Engineering Education},
    publisher = {Tempus Publications},
    author = {Richmond, Sally Sue and Defranco, Joanna F. and Jablokow, Kathryn},
    year = {2014},
    pages = {1072--1082},
}

@article{kinchin_how_2000,
    title = {How a qualitative approach to concept map analysis can be used to aid learning by illustrating patterns of conceptual development},
    volume = {42},
    issn = {0013-1881, 1469-5847},
doi = {10.1080/001318800363908},
    language = {en},
    number = {1},
    urldate = {2025-02-10},
    journal = {Educational Research},
    author = {Kinchin, Ian M. and Hay, David B. and Adams, Alan},
    month = jan,
    year = {2000},
    pages = {43--57},
}

@article{gao_clip-adapter_2024,
    title = {{CLIP}-{Adapter}: {Better} {Vision}-{Language} {Models} with {Feature} {Adapters}},
    volume = {132},
    issn = {1573-1405},
    shorttitle = {{CLIP}-{Adapter}},
doi = {10.1007/s11263-023-01891-x},
    language = {en},
    number = {2},
    urldate = {2025-09-09},
    journal = {International Journal of Computer Vision},
    author = {Gao, Peng and Geng, Shijie and Zhang, Renrui and Ma, Teli and Fang, Rongyao and Zhang, Yongfeng and Li, Hongsheng and Qiao, Yu},
    month = feb,
    year = {2024},
    pages = {581--595},
}

@article{mcinnes_umap_2018,
    title = {{UMAP}: {Uniform} {Manifold} {Approximation} and {Projection}},
    volume = {3},
    number = {29},
    journal = {The Journal of Open Source Software},
    author = {McInnes, Leland and Healy, John and Saul, Nathaniel and Grossberger, Lukas},
    year = {2018},
    pages = {861},
}

@book{tukey_exploratory_1977,
    address = {Reading, Mass},
    series = {Addison-{Wesley} series in behavioral science},
    title = {Exploratory data analysis},
    isbn = {978-0-201-07616-5},
    publisher = {Addison-Wesley Pub. Co},
    author = {Tukey, John W.},
    year = {1977},
}

@InProceedings{pmlr-v133-desmond21a,
  address = {Virtual},
  title = 	 {Semi-Automated Data Labeling},
  author =       {Desmond, Michael and Duesterwald, Evelyn and Brimijoin, Kristina and Brachman, Michelle and Pan, Qian},
  booktitle = 	 {Proceedings of the NeurIPS 2020 Competition and Demonstration Track},
  pages = 	 {156--169},
  year = 	 {2021},
  editor = 	 {Escalante, Hugo Jair and Hofmann, Katja},
  volume = 	 {133},
  series = 	 {Proceedings of Machine Learning Research},
  month = 	 {06--12 Dec},
  publisher =    {PMLR},
  url = 	 {https://proceedings.mlr.press/v133/desmond21a.html}
}

@article{hoseini2024survey,
  title={A survey on semantic data management as intersection of ontology-based data access, semantic modeling and data lakes},
  author={Hoseini, Sayed and Theissen-Lipp, Johannes and Quix, Christoph},
  journal={Journal of Web Semantics},
  volume={81},
  pages={100819},
  year={2024},
  publisher={Elsevier}
}

@article{jing2024fineclip,
  title={Fineclip: Self-distilled region-based clip for better fine-grained understanding},
  author={Jing, Dong and He, Xiaolong and Luo, Yutian and Fei, Nanyi and Yang, Guoxing and Wei, Wei and Zhao, Huiwen and Lu, Zhiwu},
  journal={Advances in Neural Information Processing Systems},
  volume={37},
  pages={27896--27918},
  year={2024}
}

@misc{monsefi_detailclip_2025,
	title = {{DetailCLIP}: {Detail}-{Oriented} {CLIP} for {Fine}-{Grained} {Tasks}},
	shorttitle = {{DetailCLIP}},
	url = {http://arxiv.org/abs/2409.06809},
	doi = {10.48550/arXiv.2409.06809},
	urldate = {2026-04-21},
	publisher = {arXiv},
	author = {Monsefi, Amin Karimi and Sailaja, Kishore Prakash and Alilooee, Ali and Lim, Ser-Nam and Ramnath, Rajiv},
	month = mar,
	year = {2025},
	note = {arXiv:2409.06809 [cs]},
}

@article{klein_critical_1989,
    title = {Critical decision method for eliciting knowledge},
    volume = {19},
    issn = {2168-2909},
    url = {https://ieeexplore.ieee.org/abstract/document/31053},
    doi = {10.1109/21.31053},
    number = {3},
    urldate = {2026-04-22},
    journal = {IEEE Transactions on Systems, Man, and Cybernetics},
    author = {Klein, G.A. and Calderwood, R. and MacGregor, D.},
    month = may,
    year = {1989},
    pages = {462--472},
}

@article{wang_defining_2009,
    title = {Defining and applying knowledge conversion processes to a visual analytics system},
    volume = {33},
    issn = {0097-8493},
    url = {https://www.sciencedirect.com/science/article/pii/S0097849309000909},
    doi = {10.1016/j.cag.2009.06.004},
    number = {5},
    urldate = {2026-04-22},
    journal = {Computers \& Graphics},
    author = {Wang, Xiaoyu and Jeong, Dong Hyun and Dou, Wenwen and Lee, Seok-Won and Ribarsky, William and Chang, Remco},
    month = oct,
    year = {2009},
    pages = {616--623},
}

@article{gavrilova_knowledge_2012,
    title = {Knowledge elicitation techniques in a knowledge management context},
    volume = {16},
    issn = {1367-3270},
    url = {https://doi.org/10.1108/13673271211246112},
    doi = {10.1108/13673271211246112},
    number = {4},
    urldate = {2026-04-22},
    journal = {Journal of Knowledge Management},
    author = {Gavrilova, Tatiana and Andreeva, Tatiana},
    editor = {Schiuma, Giovanni},
    month = jul,
    year = {2012},
    pages = {523--537},
}

@techreport{lacey2009qualitative,
  author       = {Lacey, Anne and Luff, Donna},
  title        = {Qualitative Data Analysis},
  institution  = {NIHR Research Design Service for the East Midlands and Yorkshire \& the Humber},
  address      = {Sheffield, UK},
  year         = {2009},
}

@inproceedings{rietz_cody_2021,
	address = {New York, NY, USA},
	series = {{CHI} '21},
	title = {Cody: {An} {AI}-{Based} {System} to {Semi}-{Automate} {Coding} for {Qualitative} {Research}},
	isbn = {978-1-4503-8096-6},
	shorttitle = {Cody},
	url = {https://dl.acm.org/doi/10.1145/3411764.3445591},
	doi = {10.1145/3411764.3445591},
	urldate = {2026-04-23},
	booktitle = {Proceedings of the 2021 {CHI} {Conference} on {Human} {Factors} in {Computing} {Systems}},
	publisher = {Association for Computing Machinery},
	author = {Rietz, Tim and Maedche, Alexander},
	month = may,
	year = {2021},
	pages = {1--14},
}

@article{jiang_supporting_2021,
	title = {Supporting {Serendipity}: {Opportunities} and {Challenges} for {Human}-{AI} {Collaboration} in {Qualitative} {Analysis}},
	volume = {5},
	shorttitle = {Supporting {Serendipity}},
	url = {https://dl.acm.org/doi/10.1145/3449168},
	doi = {10.1145/3449168},
	number = {CSCW1},
	urldate = {2026-04-23},
	journal = {Proc. ACM Hum.-Comput. Interact.},
	author = {Jiang, Jialun Aaron and Wade, Kandrea and Fiesler, Casey and Brubaker, Jed R.},
	month = apr,
	year = {2021},
	pages = {94:1--94:23},
}

@article{espadoto_toward_2021,
    title = {Toward a {Quantitative} {Survey} of {Dimension} {Reduction} {Techniques}},
    volume = {27},
    issn = {1941-0506},
    url = {https://ieeexplore.ieee.org/abstract/document/8851280},
    doi = {10.1109/TVCG.2019.2944182},
    number = {3},
    urldate = {2026-07-10},
    journal = {IEEE Transactions on Visualization and Computer Graphics},
    author = {Espadoto, Mateus and Martins, Rafael M. and Kerren, Andreas and Hirata, Nina S. T. and Telea, Alexandru C.},
    month = mar,
    year = {2021},
    pages = {2153--2173},
}

@article{becht_dimensionality_2019,
    title = {Dimensionality reduction for visualizing single-cell data using {UMAP}},
    volume = {37},
    copyright = {2018 Springer Nature America, Inc.},
    issn = {1546-1696},
    url = {https://www.nature.com/articles/nbt.4314},
    doi = {10.1038/nbt.4314},
    language = {en},
    number = {1},
    urldate = {2026-07-10},
    journal = {Nature Biotechnology},
    publisher = {Nature Publishing Group},
    author = {Becht, Etienne and McInnes, Leland and Healy, John and Dutertre, Charles-Antoine and Kwok, Immanuel W. H. and Ng, Lai Guan and Ginhoux, Florent and Newell, Evan W.},
    month = jan,
    year = {2019},
    pages = {38--44},
}

@article{leech_typology_2009,
    title = {A typology of mixed methods research designs},
    volume = {43},
    issn = {1573-7845},
    url = {https://doi.org/10.1007/s11135-007-9105-3},
    doi = {10.1007/s11135-007-9105-3},
    language = {en},
    number = {2},
    urldate = {2026-07-01},
    journal = {Quality \& Quantity},
    author = {Leech, Nancy L. and Onwuegbuzie, Anthony J.},
    month = mar,
    year = {2009},
    pages = {265--275},
}

@article{braun_reflecting_2019,
    title = {Reflecting on reflexive thematic analysis},
    volume = {11},
    issn = {2159-676X},
    url = {https://doi.org/10.1080/2159676X.2019.1628806},
    doi = {10.1080/2159676X.2019.1628806},
    number = {4},
    urldate = {2026-07-14},
    journal = {Qualitative Research in Sport, Exercise and Health},
    publisher = {Routledge},
    author = {Braun, Virginia and Clarke, Victoria},
    month = aug,
    year = {2019},
    note = {\_eprint: https://doi.org/10.1080/2159676X.2019.1628806},
    pages = {589--597},
}

@article{gabeff_wildclip_2024,
    title = {{WildCLIP}: {Scene} and {Animal} {Attribute} {Retrieval} from {Camera} {Trap} {Data} with {Domain}-{Adapted} {Vision}-{Language} {Models}},
    volume = {132},
    issn = {1573-1405},
    shorttitle = {{WildCLIP}},
    url = {https://doi.org/10.1007/s11263-024-02026-6},
    doi = {10.1007/s11263-024-02026-6},
    language = {en},
    number = {9},
    urldate = {2026-07-16},
    journal = {International Journal of Computer Vision},
    author = {Gabeff, Valentin and Rußwurm, Marc and Tuia, Devis and Mathis, Alexander},
    month = sep,
    year = {2024},
    pages = {3770--3786},
}

@article{kristensen-mclachlan_are_2025,
    title = {Are chatbots reliable text annotators? {Sometimes}},
    volume = {4},
    issn = {2752-6542},
    shorttitle = {Are chatbots reliable text annotators?},
    doi = {10.1093/pnasnexus/pgaf069},
    language = {eng},
    number = {4},
    journal = {PNAS nexus},
    author = {Kristensen-McLachlan, Ross Deans and Canavan, Miceal and Kárdos, Marton and Jacobsen, Mia and Aarøe, Lene},
    month = apr,
    year = {2025},
    pages = {pgaf069},
}

@misc{tableau,
  author       = {{Tableau Software}},
  title        = {Tableau},
  year         = {2026},
  howpublished = {\url{https://www.tableau.com/}},
}

\end{document}